\documentclass[nofootinbib,twocolumn]{revtex4}
\usepackage[T1]{fontenc}
\usepackage{ae,aecompl}
\usepackage{graphicx}
\usepackage{amsmath} 
\usepackage{amssymb}
\usepackage{bm}		
\usepackage{pdflscape}	
\usepackage{verbatim} 
\usepackage{mathrsfs}
\usepackage{amsfonts}
\usepackage{color}
\usepackage{hyperref}

\newcommand{\mI}{\mathcal{I}}
\newcommand{\nn}{\nonumber \\}

\newcommand{\eidi}{\epsilon_{1 \diamond}}
\newcommand{\eiidi}{\epsilon_{2 \diamond}}

  \newcommand{\ini}{\text{ini}}
  
 \newcommand{\fin}{\text{end}}
 \newcommand{\reh}{\text{reh}}
 \newcommand{\rad}{\text{rad}}
 \newcommand{\DN}{  \Delta N}  

 \graphicspath{ {../../} }

\begin{document}
\title[Inflation in non-conservative unimodular gravity]{Reconstruction of inflationary scenarios in non-conservative  unimodular gravity}

\author{Mar\'{\i}a P\'{\i}a Piccirilli}
\email{mpp@fcaglp.unlp.edu.ar}
\affiliation{Grupo de Cosmolog\'{\i}a, Facultad
	de Ciencias Astron\'{o}micas y Geof\'{\i}sicas, Universidad Nacional de La Plata, Paseo del Bosque S/N 1900 La Plata, Argentina.}
\affiliation{CONICET, Godoy Cruz 2290, 1425 Ciudad Aut\'onoma de Buenos Aires, Argentina. } 

\author{Gabriel Le\'{o}n}
\affiliation{Grupo de Cosmolog\'{\i}a, Facultad
	de Ciencias Astron\'{o}micas y Geof\'{\i}sicas, Universidad Nacional de La
	Plata, Paseo del Bosque S/N 1900 La Plata, Argentina.\\
	CONICET, Godoy Cruz 2290, 1425 Ciudad Aut\'onoma de Buenos Aires, Argentina. }

\begin{abstract}

Unimodular gravity is an alternative theory of gravity to general relativity.  The gravitational field equations are given by the trace-free version of Einstein's field equations.  Due to the structure of the theory, unimodular gravity admits a diffusion term that characterizes a possible non-conservation of the canonical energy-momentum tensor locally.  Employing this feature of unimodular gravity, in the present work, we explicitly show how to construct an inflationary phase  that can be contrasted with current observations.  In particular, we focus on three different inflationary scenarios of physical interest.   An important element in these scenarios is that the accelerated expansion is driven by the diffusion term exclusively, i.e. there is no inflaton.  Furthermore, the primordial spectrum during inflation is generated by considering inhomogeneous perturbations associated to standard hydrodynamical matter (modeled as a single ultra-relativistic fluid). For each of the scenarios, we obtain the prediction for the primordial spectrum and contrast it with recent observational bounds. 


\end{abstract}

\maketitle

\section{Introduction}\label{Sec_intro}

The inflationary paradigm, which assumes an accelerated expansion of the early Universe, is consistent with the most recent data, as reported e.g. by \textit{Planck} collaboration \cite{Planck18a,Planck18b,Planck18c}.   In the traditional version of inflation, which is also the simplest one,  the matter causing the inflationary expansion is characterized by a single scalar field--the inflaton--with a canonical kinetic term minimally coupled to gravity  described by General Relativity (GR) \cite{Guth81,Hawking82,Linde82,Linde83,Mukhanov81,mukhanov92}.   In this  simple version of inflation,  the  exponential expansion magnifies the vacuum fluctuations of the inflaton, and through some mechanism,\footnote{The specific mechanism that  describes the so called quantum-to-classical transition of the primordial perturbations  is currently a topic of  debate, the interested reader may consult Refs. \cite{Martin12,Das13,Susana13,Ashtekar2020,Leon2020,Leon2021,Berjon21,Bassi2021} and references therein. } the amplification converts them into classical perturbations, leaving their imprint as temperature and polarization anisotropies in the Cosmic Microwave Background (CMB).  

The observational data from  CMB spectrum constrain the parameters associated to the spectrum of the primordial perturbations.  In particular, \textit{Planck} 2018 collaboration \cite{Planck18b} reports the values: $\ln(10^{10} As) = 3.044 \pm 0.014 $ and $n_s = 0.9649 \pm 0.0042$ at 68\%  CL (including TT,TE,EE+lowE+lensing data). These parameters corresponds to the scalar amplitude $A_s$ and scalar spectral index $n_s$, which characterize the amplitude and shape of the primordial spectrum respectively. Another important inflationary parameter is the tensor-to-scalar ratio $r$; the \textit{Planck} 2018 collaboration reports $r < 0.061$  at 95\% CL, (using Planck TT,TE,EE+lowE+lensing+BK15 data \cite{Planck18b}), which measures the amplitude of the primordial gravitational waves that might be generated during inflation.  Given observational constraints on the inflationary parameters, namely $A_s, n_s$ and $r$, one can discriminate among different inflationary models. In particular, in the simplest version of inflation, also known as slow-roll inflation, the potential energy of the inflaton must dominate over its kinetic energy during a sufficient period.  Therefore,  the observational constraints on the inflationary parameters determine the specific inflationary potentials consistent with the data \cite{jmartinpotentials}.  

In spite of the successful predictions, there are some unsettled issues in the standard picture provided by slow-roll inflation  \cite{Ijjas2013,penrose2016fashion,Gibbons2006}.  Some of them are:  the special initial conditions required for inflation to actually begin (see however \cite{jmartin2019});    \textit{eternal inflation}, an element that is present in practically every model of inflation \cite{Kinney2016,Kinney2018,Leon17}, which leads to the controversial subject of the multiverse;  and the trans-Planckian problem for primordial perturbations \cite{jmartin2000}.   On the other hand, the precise nature and evolution of the inflaton is highly atypical, e.g. it requires negative pressure. Also, as the universe expands, the potential energy of the inflaton decreases, and eventually the  field decays into the normal particles of the standard model (and possibly into dark matter particles),  this sounds remarkably similar to the Higgs mechanism, except that the inflaton, in its simple formulation, cannot be the Higgs field.\footnote{ Nonetheless, if the inflaton is non-minimally coupled to gravity, the Higgs field could lead to an inflationary expansion of the slow-roll type \cite{Bezrukov2013}.} The aforementioned process, known as reheating, is actually poorly understood from the point of view of particle physics.

In view of the previous arguments, it is reasonable to explore alternatives to the traditional slow-roll inflationary model, which may account for some of the problems mentioned but leave the successful predictions of inflation unchanged. Unimodular Gravity (UG) is an alternative approach to GR that  can be derived from the Einstein-Hilbert action by restricting to variations preserving the volume element \cite{UG0,UG1,UG2,UG3,UG4,ellis2010}. The application of UG into the cosmological context has been noted for decades \cite{Weinberg89,ellis2010,UG4,Smolin2009,Uzan2010} (in particular with its connection to the cosmological constant), but recently it has been rediscovered, gaining significant attention \cite{ellis2015,Nojiri2015,Nojiri2016,Corral20,Nucamendi20,sudarskyPRL1,sudarskyPRL2,sudarskyH0,Daouda2018,miguelangel1,miguelangel2,GUG1,GUG2,deCesare2021,Fabris2021a,Fabris2021b,aperez2021,gabrielUG}. 

The gravitational field in UG is characterized by the trace-free Einstein equations, and due the structure of the theory, UG admits a possible non-conservation of the energy-momentum tensor, represented by a \textit{diffusion term}. Namely,  the conservation of the energy-momentum tensor  is considered as an extra hypothesis that, when imposed, one recovers the original Einstein Field Equations (EFE) with a cosmological constant, which is now an integration constant \cite{Finkelstein1971,Weinberg89,ellis2010,ellis2015}.   Therefore, all the successful predictions of GR are included in the UG theory.  However, one can choose not to impose the energy-momentum conservation; this approach leads to non-conservative UG.\footnote{We invite the reader to consult Refs. \cite{Elias2020,Hermano2021}, for a review of proposals discussing the theoretical and conceptual constructions surrounding the non-conservation of the energy-momentum tensor.}

 In Refs.  \cite{sudarskyPRL1,sudarskyPRL2,sudarskyH0}, the energy-momentum non-conservation was argued to arise due to a fundamental granularity of the spacetime at Planckian scales. Moreover,  in Ref. \cite{aperez2021} that same approach was used to characterize an inflationary phase.  In that work, the primordial inhomogeneities, born at the Planck scale, are generated  by  the interaction of the homogeneous part of the
 Higgs field and the inhomogeneous granular structure at
 the Planck scale  (mediated by the scale invariance-breaking of the Higgs); it is also important to mention that  this mechanism is based on the semiclassical gravity framework.  However, the inflationary expansion was driven exclusively by the diffusion term, i.e. there was no inflaton. 
 
 In Ref. \cite{gabrielUG},  we analyzed the generic case of a quasi-de Sitter expansion due to the diffusion term, i.e. without specifying the details of the microphysics at stake.  In particular, in that work we derived the  generic conditions required for any type of diffusion to generate a realistic  inflationary epoch (so the particular model of Ref. \cite{aperez2021} could be included in that analysis).   Also, for a given parameterization of inflation in terms of the Hubble flow functions (HFF) \cite{terreroHFF,terreroHFF2}, we showed how to reconstruct the corresponding diffusion term in such a way that a smooth transition occurs between inflation and the subsequent radiation dominated era, hence reheating proceeds in a natural manner.  The primordial spectrum was produced by assuming inhomogeneous perturbations corresponding to a single ultra-relativistic fluid, that is we considered standard hydrodynamical matter.\footnote{Additionally,  in Ref. \cite{gabrielUG}, we analyzed the feasibility of identifying  the diffusion term, responsible for the inflationary expansion,  with the current observed value of the cosmological constant.}

 In the present paper, based on the results of Ref.  \cite{gabrielUG}, we will perform a phenomenological analysis.  Specifically, we will explore three different inflationary scenarios in non-conservative UG.  The first scenario is motivated by its simplicity, namely we choose the most simple parameterization of inflation in terms of the HFF.  For the second scenario, we choose a parameterization of the diffusion term which was originally introduced in \cite{gabrielUG}. That parameterization is motivated by the fact  it can produce an inflationary phase and at the same time account for the present value of the cosmological constant. In addition, this same parameterization can accommodate, as a particular case, a model equivalent to that of Ref. \cite{aperez2021}.  Lastly, the third scenario consists of  parameterizing any slow-roll inflation model through the HFF. Using these HFF, we find its corresponding non-conservative UG version (hence the third scenario actually consists of multiple possible cases).  In this manner, we can argue that any single field inflationary model of the slow-roll type can be mapped into an inflationary model in non-conservative UG. For each of the three scenarios described previously, we  reconstruct the corresponding  diffusion term. Afterwards,  we compute the theoretical predictions for the inflationary parameters $A_s$, $n_s$ and $r$.  Thus, we can find a range of values for the  parameters of the diffusion term that are consistent with the data. Our phenomenological analysis is in essence equivalent to the traditional method of constraining the parameters of slow-roll inflation (for instance as presented in \cite{jmartinpotentials,Planck18b}).  This is, using observational bounds, one is able to constrain the parameters of a given diffusion term (or potential in slow-roll inflation) that might lead to inflation but without specifying in great detail the high energy theory that leads to a specific diffusion term (or potential  in slow-roll inflation).

The paper is organized as follows: In Sec.  \ref{Sec_cosmoUG}, we provide a very brief review of UG, focusing on its implementation into the cosmological context. We also introduce the conditions required for a generic diffusion term to generate an inflationary phase. Additionally, we show how to reconstruct the diffusion term through the HFF. Also, we present a preliminary analysis on how to include scalar fields in our proposal. In Sec. \ref{Sec_theory},  we explicitly find the diffusion term for each of the three scenarios considered, and also present the theoretical predictions that will be of observational interest.  In Sec. \ref{Sec_observations},  we present the results of our analysis using the observational data, and exhibit the empirical constraints on the parameters of each scenario. Finally, in Sec. \ref{Sec_conclusions}, we summarize the main results of the paper and present our conclusions. Throughout this work, we will use a $(-,+,+,+)$ signature for the spacetime metric and units where $c=1=\hbar$.

\section{Cosmological background in non-conservative unimodular gravity }\label{Sec_cosmoUG}
In this section we will introduce the cosmological equations in non-conservative UG. In particular,  we will focus on presenting the mechanism responsible to produce an inflationary scenario without inflaton.  A more detailed presentation can be found in \cite{gabrielUG}.

The UG action  can be expressed through the functional
\begin{eqnarray}\label{accion0}
	S[g^{ab}, \Psi_M;\lambda] &=& \frac{1}{2 \kappa} \int \left[  R \epsilon_{abcd}^{(g)}  -2 \lambda (\epsilon_{abcd}^{(g)}  - \varepsilon_{abcd} ) \right]  \nn
	&+& S_M [g^{ab}, \Psi_M] ,
\end{eqnarray}
where $\kappa \equiv 8 \pi G$, $\epsilon_{abcd}^{(g)} $ is the 4-volume element associated to the metric $g_{ab}$, $\varepsilon_{abcd} $ is a fiduciary 4-volume element (which we take as a given) and  $R$ is the Ricci scalar. The scalar $\lambda(x)$ is a Lagrange multiplier function, and  $S_M$ is the action of the matter fields represented collectively by $\Psi_M$. 

Invoking the variational principle and applying it to action \eqref{accion0},   results in the equations of motion. Specifically, the equations of motion are obtained by requiring $\delta S =0$, and performing variations of Eq.  \eqref{accion0} with respect to the dynamical variables: $g^{ab}$, $\lambda$ and $\Psi_M$; this procedure yields 
\begin{equation}\label{EFE}
	R_{ab} - \frac{R}{2}g_{ab} + \lambda(x) g_{ab} = \kappa T_{ab},
\end{equation}
\begin{equation}\label{constraint0}
	\epsilon_{abcd}^{(g)}  = \varepsilon_{abcd},
\end{equation}
\begin{equation}\label{KG}
	\frac{\delta S_M}{\delta \Psi_M} = 0, 
\end{equation}
where 
\begin{equation}\label{defTab}
	T_{ab} \equiv  \frac{-2}{\sqrt{-g}}  \frac{\delta S_M}{\delta g^{ab}}
\end{equation}
is the energy-momentum tensor. Equation \eqref{KG} is a Klein-Gordon type of equation for the matter fields. Also $g$ denotes the determinant of the components of the metric tensor $g_{\mu \nu}$ (in a specific coordinate basis). 

As  can be seen directly from  \ref{EFE},  taking the covariant derivative on both sides of the equation shows that in general $\nabla^a T_{ab} \neq 0$.  In other words, UG generically admits a violation of the energy-momentum conservation, thus the name non-conservative UG.  

In fact, it is clear that action \eqref{accion0}--as a whole--is invariant under generic one-parameter family of diffeomorphisms, because it is constructed from an integral of geometrical objects that are defined on a manifold.  In particular, we can consider the variation of action \eqref{accion0} with respect to all geometrical objects, i.e.  by considering diffeomorphisms acting  on the dynamical variables ($g^{ab}$, $\lambda$ and $\Psi_M$), and also on the non-dynamical ones ($\varepsilon_{abcd}$); this procedure results in $\delta S =0$ directly.  Using that property, together with the equations of motion, it can be shown \cite{sudarsky2023} that  
\begin{equation}\label{conservEM}
	\nabla^a (T_{a b} - g_{a b} Q) = 0,
\end{equation}
where $Q(x)$ is  an arbitrary function that quantifies the violation of the conservation of  $T_{ab}$, the scalar $Q$ will be referred to as the \textit{diffusion term}.  
Equation \eqref{EFE} along with with Eq. \eqref{conservEM}, leads to $\nabla_ a \lambda(x) =  \kappa \nabla_a  Q(x)$, which can be solved as
\begin{equation}\label{lambdaeff}
	\lambda(x) = \Lambda_* + \kappa Q
\end{equation}
In this case $\Lambda_*$ is an integration constant fixed by the initial data. On the other hand, if one fixes $Q=$ constant, the usual conservation law is recovered $\nabla^a T_{a b} = 0 $. Therefore, conservation of $T_{ab}$ can be introduced in UG as an additional premise, bringing us back to GR (with a cosmological constant given by $\Lambda_*$).  In this work, we will not consider such possibility, and focus on the non-conservative version of UG.

We can eliminate the Lagrange multiplier from Eq. \eqref{EFE}.   Taking the trace of such an equation results in 
\begin{equation}\label{lambda}
	\lambda = \frac{1}{4} \left( \kappa T + R \right),
\end{equation}
where $T = g^{ab} T_{ab}$ is the trace of the energy-momentum tensor.  Substituting the former expression in Eq. \eqref{EFE},  leads to the trace-free part of Einstein's field equations, namely
\begin{equation}\label{UGecs}
	R_{ab} - \frac{1}{4}  g_{ab} R  =   \kappa \left( T_{ab} - \frac{1}{4}  g_{ab} T \right).
\end{equation}
These are UG equations for the gravitational field.

Assuming the cosmological principle and spatial flatness, the line element is given by the Friedmann-Lemaitre-Robertson-Walker (FLRW) metric
\begin{equation}\label{metricaFRW}
	ds^2 = -dt^2  + a^2 (t) (dx^2 + dy^2 + dz^2).
\end{equation}
From the previous metric, we can establish that the fiduciary volume element  is
\begin{equation}\label{constraintFRW}
	\varepsilon_{abcd}= \epsilon^{(g)}_{abcd} = a^3 e_{abcd},
\end{equation}
where
\begin{equation}\label{eabcd}
	e_{abcd} = dt_a \wedge dx_b \wedge dy_c \wedge dz_d.
\end{equation}
or in shorthand notation $d^4x$.

The matter content in the universe is modeled as a perfect fluid, namely the energy-momentum tensor is
\begin{equation}\label{EMtensor}
	T_{\mu \nu} = (\rho +P ) u_\mu u_\nu + P g_{\mu \nu},
\end{equation}
where $\rho$ and $P$ are the energy density and pressure in the rest frame of the fluid respectively, and $u^\mu$  represents its 4-velocity (relative to the observer) normalized as $u^\mu u_\mu = -1 $.  Furthermore,  we assume that the energy-momentum tensor and the diffusion term $Q$ are also spatially  isotropic and homogeneous, i.e. $\rho(t)$, $P(t)$ and $Q(t)$.  The resulting set of Friedmann's equations is thus
\begin{equation}\label{F1}
	3	H^2 = \frac{1}{ M_P^2} (\rho + Q) ,
\end{equation}
\begin{equation}\label{F2}
	2 \dot H + 3 H^2 = \frac{1}{M_P^2} (-P + Q),
\end{equation}
where $M_P^2 = \kappa^{-1}$ is the reduced Planck mass.  The dot denotes derivative with respect to cosmic time $t$ and $H \equiv \dot a/a$ is the Hubble factor.  Note that we have set $\Lambda_* =0$ since we are always free to choose the initial values $H^\ini$, $\rho^\ini$ and $Q^\ini$ in such a way that the integration constant vanishes. 

 The  non-conservation of the energy-momentum tensor, Eq. \eqref{conservEM}, implies
\begin{equation}\label{rhomov}
	\dot \rho + \dot Q + 3 H(\rho + P) = 0.
\end{equation}
Thus, given  a specific form of the diffusion term  $Q(t)$ and an equation of state $P(\rho)$, the system of equations is closed.

Another important aspect in the theory that becomes modified by the non-conservation of the canonical energy-momentum tensor, involves the so called \textit{energy conditions} associated to $T_{ab}$. In fact, this subject has been investigated before in Extended Theories of Gravity, and in particular, when considering non-conservation of the energy-momentum tensor, see e.g. \cite{Hermano2021,Capozziello2014}.

In order to analyze such a feature, it is convenient to rewrite the UG equations by substituting Eq. \eqref{lambdaeff} in \eqref{EFE}, obtaining
\begin{equation}\label{EFE2}
R_{ab} - \frac{R}{2}g_{ab} + \kappa Q g_{ab} = \kappa T_{ab},
\end{equation}
where we have considered that $\Lambda_* =0$ (without loss of generality). A common practice is to move the term $\kappa Q g_{ab}$ to the r.h.s of \eqref{EFE2}, and define $T_{ab}^{\textrm{eff}} \equiv T_{ab} - Q g_{ab}$, i.e. to define an effective energy-momentum tensor. Then, the energy conditions are attributed to $T_{ab}^{\textrm{eff}}$. However, as explained in \cite{Capozziello2014}, this procedure hides the energy conditions of the real matter fields, characterized by $T_{ab}$, because it combines the matter degrees of freedom with a geometrical object $Q g_{ab}$.\footnote{The non-conservation equation $\nabla_a T^{ab} = g^{ab} \nabla_a Q$ arises because of restricting the theory to diffeomorphisms that preserve the 4-volume element (see constraint \eqref{constraint0}); so $Q g_{ab}$ is of a geometric nature. } Hence, to analyze the energy conditions in non-conservative UG we proceed as follows.

The \textit{weak energy condition} (WEC) is defined as
\begin{equation}\label{wec}
	T_{ab} X^a X^b \geq 0
\end{equation}
where $X^a$ is a timelike vector, i.e. $g_{ab} X^a X^b = -1$. Substituting Eq. \eqref{EFE2} in \eqref{wec}, and taking into account Eq. \eqref{lambda}, yields
\begin{equation}\label{wec2}
	R_{ab} X^a X^b \geq \kappa \left( \frac{T}{2} - Q \right).
\end{equation}
On the other hand, the WEC in standard GR translates into  $R_{ab} X^a X^b \geq \kappa T/2$. Therefore, the second term in the right-hand side of Eq. \eqref{wec2} is a modification to the usual condition on the geometrical objects associated to the WEC.

The \textit{strong energy condition} (SEC) is defined as
\begin{equation}\label{sec}
	 \left( T_{ab} - \frac{T}{2}  g_{ab} \right)  X^a X^b \geq 0,
\end{equation}
where $X^a$ a timelike vector. In standard GR, and through the traditional EFE, the above condition takes the form $R_{ab} X^a X^b \geq 0$. However, in UG by substituting Eqs. \eqref{EFE2} and \eqref{lambda} in \eqref{sec}, we obtain 
\begin{equation}\label{sec2}
	R_{ab} X^a X^b \geq -\kappa Q.
\end{equation}
We note that if $Q>0$ and if $R_{ab} X^a X^b$ satisfies the corresponding WEC and SEC inequalities in standard GR, then $R_{ab} X^a X^b$ automatically satisfies Eqs. \eqref{wec2} and \eqref{sec2}, i.e. the WEC and SEC inequalities of UG. 

Here it is important to mention that the so-called \textit{Raychaudhuri Equation}, which is obtained from the geometrical term $R_{ab} X^a X^b$, is modified when one considers the diffusion term of Eq. \eqref{EFE2}. In addition, the non-conservation of the enery-momentum tensor implies the non-geodesic motion of pointlike particles \cite{yuri2022}.

For completeness, the \textit{dominant energy condition}  states that for all future directed, timelike $Y^a$, the vector $-T^a_b Y^b$ should be a future directed timelike or null vector. This condition translates, through Eq. \eqref{EFE2}, that $-[R^a_b + (\kappa Q - R/2) \delta^a_b] Y^b$ should be a future directed timelike or null vector.  Finally, the \textit{null energy condition} states that $T_{ab} Z^a Z^b \geq 0$, where $Z^a$ is a null vector, i.e. $g_{ab} Z^a Z^b =0$. The NEC and Eq. \eqref{EFE2} imply that $R_{ab} Z^a Z^b \geq 0$ in UG, which is the same condition as in standard GR.


\subsection{Characterizing the inflationary phase}

The inflationary phase can be parameterized completely  by the Hubble flow functions (HFF),  \cite{terreroHFF,terreroHFF2} defined as
\begin{equation}\label{HFF}
	\epsilon_{n+1} =\frac{ \dot \epsilon_{n}}{H \epsilon_{n}}, \qquad \epsilon_{0}=\frac{H^\ini}{H},
\end{equation} 
where $n =0, 1, 2, \ldots$.  From Eqs.  \eqref{F1} \eqref{F2},  \eqref{HFF}, and   assuming the equation of state (EOS) for the perfect fluid as $P = w \rho$, one obtains,
\begin{equation}\label{eps1gen}
	\epsilon_1 =  \frac{3(1 + w)}{2(1+\Gamma)},  \qquad   \text{where}  \qquad  \Gamma \equiv Q/\rho.
\end{equation} 
We observe from  the previous equation that if $\Gamma$ satisfies  $\Gamma > (1+3w)/2$, then $\epsilon_{1} < 1$; meaning that an inflationary regime can take place.

Our particular model assumes that,  from the beginning of the inflationary expansion, the total matter content of the universe  behaves as a hydrodynamical fluid consisting of pure radiation.  
Hence, the EOS $p = w \rho$ with $w=1/3$  is valid from the beginning of inflation, during inflation, and up to near the end of the radiation dominated epoch (see \cite{gabrielUG} for more details about the post-inflationary epoch).  

For $w=1/3$, then 
\begin{equation}\label{epsilon1}
	\epsilon_1 = \frac{2}{ 1 + \Gamma}.
\end{equation}
Therefore, inflation occurs i.e. $\epsilon_1 < 1$, if
\begin{equation}\label{condicionQinf}
	\frac{Q}{\rho} > 1.
\end{equation}
In other words, if the diffusion term $Q$ dominates over $\rho$ (with EOS parameter $w=1/3$), then inflation is assured.  This means that if inflation takes place, then the accelerated expansion is being driven by $Q$ (and not by $\rho$ as in the traditional picture). 

On the other hand, inflation ends when $\epsilon_{1} \simeq 1$; consequently,  the inflationary phase ends when $Q \simeq \rho$. Given that during inflation we are modeling the matter content as  a perfect fluid consisting of pure radiation,  this means that the universe evolves into the radiation era smoothly, that is, reheating takes places naturally.

For $w=1/3$, the continuity equation \eqref{rhomov} is expressed as
\begin{equation}\label{rhomovN}
	\rho(N),_N + Q(N),_N +4 \rho(N) =0,
\end{equation}
where $ ( \cdot ),_N$ denotes derivative  with respect to  $N$. Also, we have performed a change of variable $H dt = dN$, where $N$ is the number of e-folds,  defined as $a(N) \equiv e^{N} a_\ini$.  The solution to Eq. \eqref{rhomovN} for a given generic ``initial'' condition $\rho(N_0) = \rho_0$ is\footnote{Note that the name ``initial'' condition, does not necessarily means that  one must choose $N_0 = 0$. In fact, the allowed ranges are: $0 \leq N_0 \leq N_f$, and $ M_P^4 \leq \rho_0 \leq \rho_\fin$.}
\begin{equation}\label{rhosolalt}
	\rho (N)= \rho_0 e^{-4 (N- N_0)} - e^{-4 N} \int_{N_0}^N e^{4 \bar{N}} Q,_{\bar N} d\bar N.
\end{equation}
At this point, the only thing left to specify is the function $Q$, which would yield $\rho$ from \eqref{rhosolalt}.  In particular, one could analyze whether for a given diffusion term, the condition \eqref{condicionQinf} is satisfied. If this is the case, then $Q$ would generate an inflationary phase up to $Q \simeq \rho$, where the end of inflation would occur after sufficient $N$.

Alternatively,  we can reconstruct a specific form of $Q$ such that is compatible with a realistic inflationary phase. Our strategy is as follows:  we will assume a particular function $\epsilon_1$ that is compatible with a full inflationary phase.  For that chosen $\epsilon_{1}$, we construct the corresponding $Q$,  and  consider it as an ansatz.   In this manner, we can analyze some observational consequences for that constructed $Q$ (for a more in depth analysis see Ref. \cite{gabrielUG}). 

Generalizing the analysis of Ref. \cite{gabrielUG}, for a given $\epsilon_{1}$,  we have 
\begin{equation}\label{rhosolNgen}
	\rho(N)= \rho_0 \frac{\epsilon_1(N)}{\epsilon_{1}(N_0)} \exp[-  \int_{N_0}^N 2 \epsilon_1 (\bar N)   d\bar N  ].
\end{equation}
and
\begin{equation}\label{QNgen}
	Q(N) = Q_0  \left(  \frac{2-\epsilon_{1} (N)}{2-\epsilon_{1}(N_0)}  \right) \exp[-  \int_{N_0}^N 2 \epsilon_1 (\bar N)   d\bar N  ]. 
\end{equation}
Note however that $Q_0$ and $\rho_0$ are not independent; in fact from \eqref{epsilon1}, we have
\begin{equation}\label{relacionQ0rho0}
	Q_0 = \left(  \frac{2}{\epsilon_1 (N_0) } - 1  \right) \rho_0.
\end{equation}

It can be shown that Eqs. \eqref{rhosolNgen} ,  \eqref{QNgen}, consistently solve Eq. \eqref{rhomovN}.  Thus, we have inverted the problem, for any inflationary phase completely characterized by $\epsilon_{1}(N)$, we can find the corresponding diffusion term, and thus the evolution of the  energy density during the accelerated expansion.

\subsection{Possible options for including scalar fields}

Before ending this section, we would like to discuss two possible manners of introducing scalar fields in the proposed model. In this way, we hope to inspire some model building prospects within such a framework.

First, we can consider the following  toy model: we maintain the assumption that $Q$ is a positive diffusion term, but characterize the matter fields by a scalar field $\phi(x)$ minimally coupled to gravity. The scalar field is the only type of matter (or dominates over other matter degrees of freedom). Therefore, from Eq. \eqref{defTab}, we have that
\begin{equation}\label{Tabphi}
	T_{ab} = \nabla_a \phi \nabla_b \phi +  \left[ X  - V(\phi) \right] g_{ab}, 
\end{equation}
where we have defined the kinetic term as
\begin{equation}
	X \equiv - \frac{1}{2} g^{ab} \nabla_a \phi \nabla_b \phi.
\end{equation}
The non-conservation equation of $T_{ab}$ \eqref{conservEM} leads to
\begin{equation}\label{conservphi}
 (\nabla^a \nabla_a \phi - \partial_\phi V ) \nabla_b \phi = \nabla_b Q.
\end{equation}

The energy momentum tensor in Eq. \eqref{Tabphi}, can be rewritten as characterizing a ``perfect fluid'' for the scalar field, i.e.
\begin{equation}
	T_{ab} = (\rho_\phi + P_\phi)U_a U_b + P_\phi g_{ab},
\end{equation}
where the energy density and pressure associated to the scalar field are given as
\begin{equation}\label{rhoPphi}
	\rho_\phi = X + V, \qquad P_\phi= X-V,
\end{equation}
with 4-velocity $U^a = -g^{ab} \nabla_b \phi/\sqrt{2X}$. 

We will now assume that the spacetime is homogeneous/isotropic, and, for simplicity, spatially flat. Therefore, the spacetime is described by the flat FLRW metric. On the other hand, we can also consider that the scalar field can be approximated by a homogeneous scalar field, i.e. $\phi \simeq \phi(t)$, so $X=\dot \phi^2/2$. Under these assumptions,  Eq. \eqref{eps1gen} is expressed as
\begin{equation}\label{eps1phi}
	\epsilon_1 = \frac{3}{2} \frac{(\rho+P)}{(\rho+Q)} = \frac{3X}{X+V+Q},
\end{equation}
where in the last equality we have used Eqs. \eqref{rhoPphi}. Therefore, the condition for an inflationary phase $\epsilon_{1} <1$, obtained from \eqref{eps1phi}, is
\begin{equation}\label{condinf99}
	V+Q > 2X.
\end{equation}
Thus, if the diffusion term dominates over the kinetic term $Q \gg X$, then $Q$ can drive the accelerated expansion in the early universe. This conclusion is independent if $V$ dominates or not over $X$, i.e. $\phi$ may not be the inflaton. Additionally, the inflationary phase will end when $V+Q \simeq 2X$. Evidently, the next step to analyze the feasibility of this toy model is to consider small inhomogeneous perturbations around the scalar field and the background spacetime. This study is beyond the scope of the present work, but it is an interesting possibility for future research.

The second option to introduce scalar fields, is to consider in addition a radiation fluid. Consequently, the total matter content is modeled as radiation plus a scalar field. The radiation fluid can be thought of as consisting of electromagnetic radiation (and possibly gravitons). 

As before, we describe the background spacetime as a spatially flat FLRW; so the background matter content is homogeneous and isotropic. Specifically, the energy density and pressure of the radiation fluid, $\rho_r$ and $ P_r$ respectively, are highly homogeneous. Also, we assume that the scalar field can be decomposed in a homogeneous background plus a small (inhomogeneous) perturbation, $\phi(x) = \phi_0 (t) + \delta \phi(x)$.

Additionally, we assume that the diffusion term, which could emerge due to quantum gravitational effects \cite{sudarskyPRL1,sudarskyPRL2,aperez2021}, decreases during the expansion, and feeds the radiation fluid. 
 Therefore, the continuity equation for radiation is expressed as
\begin{equation}\label{rhorad0}
	\dot \rho_r + 3H(\rho_r + P_r) = - \dot Q_0,
\end{equation}
where $P_r = \rho_r/3$. In the above equation, a key assumption is that the diffusion process  does not disrupt the homogeneity and isotropy  of the background matter and geometry configurations to leading order. In other words we are assuming $Q=Q_0 (t) + \delta Q(x)$, with $Q_0 \gg |\delta Q|$. For the background part of the scalar field one has,
\begin{equation}\label{rhophi0}
	\dot \rho_\phi + 3H(\rho_\phi + P_\phi) = 0 
\end{equation}
where $\rho_\phi$ and $P_\phi$ are given in Eq. \eqref{rhoPphi}, with $X = \dot \phi_0^2/2$ and $V= V(\phi_0)$. Consequently, the continuity equation for the total matter content of the background, i.e. Eq. \eqref{rhorad0} + Eq. \eqref{rhophi0}, reflects the non-conservation of the total background $T_{ab}$. 

Next, we will analyze the conditions for generating an accelerated expansion. If we assume that the scalar field dominates over the radiation fluid, then  condition \eqref{condinf99} is maintained, i.e. $Q_0+V > 2X$ implies an inflationary phase. Note that, as in the previous case, if the diffusion term is much larger than the kinetic term, $Q_0 \gg X$, then the exponential expansion is driven exclusively by $Q_0$. 

As we have mentioned, the origin of the diffusion term could be linked to a possible fundamental granularity of the spacetime at Planckian scales. The interaction of the scalar field with this granularity could  generate the primordial perturbations. Phenomenologically, this interaction could be represented by the term $\delta Q(x)$. Then, one could use Eq. \eqref{conservphi} at the leading order in $\delta \phi, \delta Q, \delta g_{ab}$ to analyze the behavior of the perturbations.

On the other hand,  it may well be the case that the term $\delta Q(x)$ is dominated by thermal fluctuations from the radiation fluid. In this case, the whole scenario resembles to the so called \textit{warm inflation} framework \cite{warm1,warm2,warm3,warm4}. In the latter, the  interaction of the scalar field with the heat reservoir (modeled by the radiation fluid) induces primordial perturbations. However, there are some caveats when making the association between warm inflation and our model based in non-conservative UG. Clearly, the most glaring difference is that in warm inflation the total energy-momentum tensor is conserved. The other distinction is that in warm inflation, the scalar field drives the accelerated expansion. However, this latter discrepancy can be eliminated by considering that, in our model, the potential $V$ dominates over the kinetic term, $V \gg X$, so the condition for inflation is fulfilled, i.e. $Q+V > 2X$.  In this situation, the entire framework is essentially warm inflation except for the non-conservation of the total energy-momentum tensor. In fact, this feature might be helpful for solving the grateful exit problem in warm inflation \cite{warm5}, because in our proposal inflation would end when $V+Q \simeq 2X$. Thus, much of the phenomenology of warm inflation and its extensive studies comparing to CMB data may then become useful to the inflationary models based on non-conservative UG and vice versa.

\section{Three inflationary scenarios: Theoretical Analysis}\label{Sec_theory}

In the present section we will introduce three inflationary scenarios in non-conservative UG, each one of them has different motivations that we consider of physical interest.  Our main objective here is to derive all the theoretical equations required for comparing the predictions with observations.  

In Ref. \cite{gabrielUG} it was shown in detail that the theoretical predictions of the inflationary model  considered here, which is based in non-conservative UG,  are exactly the same as in standard (single field slow-roll) inflation.   In particular,  to achieve that result, we made two key assumptions: (i)  the total matter content in the early universe behaves  as a perfect fluid consisting of pure radiation, so $T=0=\delta T$ and (ii) the quantity $Q$, characterizing the non-conservation of the energy-momentum tensor, is completely homogeneous, so $\delta R = \delta Q =0$.  Furthermore, under assumptions (i) and (ii), in \cite{gabrielUG} it was shown that the perturbed EFE at linear order in UG are exactly the same as in traditional GR, i.e. $\delta R^{\mu}_{\: \: \nu}   = \kappa \delta T^{\mu}_{\: \: \nu}$ (see also \cite{UGbranden,UGindios}).

The scalar power spectrum obtained is thus,
\begin{equation}\label{Pk0}
	P_s (k) = A_s \left(\frac{k}{k_\diamond}\right)^{n_s-1}, 
\end{equation}
where
\begin{equation}\label{paramsinf}
	A_s = \frac{H_\diamond^2}{8 \pi^2  \eidi  M_P^2},  \qquad  n_s = 1-2\eidi - \eiidi
\end{equation}
and the tensor-to-scalar ratio is
\begin{equation}\label{paramr}
	r = 16 \eidi.
\end{equation}
The $\diamond$ denotes  we are evaluating the corresponding quantity at the time of  ``horizon crossing'' for  the mode $k_\diamond$, i.e. when $k_\diamond = a_\diamond H_\diamond $. 
Consequently, if we find the corresponding expressions for: $\epsilon_{1} (N)$, $\epsilon_{2} (N)$, and $H(N)$ in the non-conservative UG framework, then it is straightforward to obtain the predicted power spectrum after evaluating such expressions at $N_\diamond$.  
Note that because of assumption (i), the mode associated to $k$ corresponds to small inhomogeneities in the matter fluid $\delta \rho_k$, consisting of pure radiation. 

Before introducing the three inflationary scenarios, it is worth mentioning another important parameter in the traditional (slow-roll) inflationary model,  i.e.  the reheating parameter \cite{jmartinreheating1,jmartinreheating2}, given as 
\begin{equation}
	\ln R_{\rad} = \frac{1-3 \bar{w}_{\reh}}{12(1+\bar{w}_{\reh})} \ln \left( \frac{\rho_{\reh}}{\rho_{\fin}} \right),
\end{equation}
where $\bar{w}_{\reh}$ is the mean equation of state parameter during reheating, $\rho_{\reh}$ is the energy density at the end of the reheating era, and $\rho_{\fin}$ is the energy density at the end of inflation.  In slow-roll inflation, the reheating parameter $R_\rad$ put constraints on the interval $N_f - N_\diamond$ \cite{jmartinreheating1,jmartinreheating2}, where $N_f$ is the total number of e-folds that inflation lasts.  However, in our inflationary model due to assumption (i),  $\bar{w}_{\reh} = 1/3$ exactly, in fact $w=1/3$ since the beginning of inflation, during inflation, after the end of inflation, and during the radiation dominated era up to the matter dominated epoch (see \cite{gabrielUG} for a more detailed presentation of the post-inflationary epoch).   As a consequence, in our model $\ln R_{\rad} = 0$, this implies that the interval $N_f - N_\diamond$ is fixed simply by the dynamical equations of the background, i.e. \eqref{rhosolNgen} and \eqref{QNgen}.  Also, it means that reheating proceeds in a smooth manner.

In the following, we will show how to obtain the main theoretical quantities involved to compute $A_s$, $n_s$ and $r$ in the three inflationary scenarios mentioned.

\subsection{First scenario}

This scenario is motivated by its simplicity. Namely, one of the most simple parameterizations of  $\epsilon_{1}$  characterizing inflation is
\begin{equation}\label{eps1caso1}
	\epsilon_1 (N) =  \frac{1}{(1+N_f-N)^\gamma}, \qquad \gamma >0,
\end{equation}
where $N_f \geq 65$, as usual. Consequently,
\begin{equation}\label{eps2caso1}
	\epsilon_2(N) =  \frac{\gamma}{1+N_f-N}.
\end{equation}
This parameterization is also inspired by the one  introduced in Ref. \cite{Mukhanov2013},  where it was argued to be one of the most general descriptions of inflation, i.e. without assuming a specific model for the matter fields other than the EOS (which in the case of Ref. \cite{Mukhanov2013} is $P \simeq -\rho$).  

From Eqs. \eqref{rhosolNgen} and \eqref{QNgen}, we obtain the following expressions:
\begin{eqnarray}\label{rhocaso1}
	\rho(N)  &=&  \frac{\rho_\fin}{(1+N_f-N)^\gamma}  \nn
	&\times& \exp \left\{      \frac{2}{1-\gamma}  \left[  \left(N_f - N + 1\right)^{1-\gamma}-1      \right]      \right\}
\end{eqnarray}
and
\begin{eqnarray}\label{Qcaso1}
		Q(N)  &=&  \left[  2 - \frac{1}{(1+N_f-N)^\gamma}    \right] \rho_\fin \nn
	&\times& \exp \left\{      \frac{2}{1-\gamma}  \left[  \left(N_f - N + 1\right)^{1-\gamma}-1      \right]      \right\}.
\end{eqnarray}
Note that we have chosen $N_0 = N_f$, so $\rho_0 = \rho_\fin$.  Then, from Friedmann equation \eqref{F1}, we have
\begin{equation}\label{Hcaso1}
	H^2(N) = \frac{2 \rho_\fin}{3 M_P^2} \exp \left\{      \frac{2}{1-\gamma}  \left[  \left(N_f - N + 1\right)^{1-\gamma}-1      \right]      \right\}
\end{equation} 
Thus, there are two parameters in this case:  $\gamma$  and $\rho_\fin$.

\subsection{Second scenario}

This second case was originally presented in Ref. \cite{gabrielUG}. The motivation there was to find a specific parameterization of $Q$ such that it can produce an inflationary phase and also coincides with the present value of the cosmological constant. That is, we explored a possibility in which inflation and the late time accelerated expansion might be unified within non-conservative UG.  In this paper, we will only analyze the observational bounds  imposed on that specific model coming from inflation. Therefore, we left for future work the same analysis employing data of the current value of the cosmological constant.

In this case, we parameterize
\begin{equation}\label{eps1caso2}
	\epsilon_1 (N) = 1 + \tanh \left[\frac{2}{3}  \alpha (N-N_f) \right] +\exp(-4 \alpha N)
\end{equation}
then
\begin{equation}\label{eps2caso2}
	\epsilon_2(N) = \frac{2 \alpha}{3} \left( \frac{-6 \exp(-4 \alpha N) + \textrm{sech}^2 \left[2/3  \alpha (N-N_f) \right] }  {1 + \exp(-4 \alpha N) + \tanh \left[2/3  \alpha (N-N_f) \right] }  \right)
\end{equation}

As we have shown in Ref. \cite{gabrielUG}, this case results in the following expressions (which can be obtained using \eqref{eps1caso2} in Eqs. \eqref{rhosolNgen}, \eqref{QNgen})
\begin{eqnarray}\label{rhocaso2}
	\frac{\rho(N)}{M_P^4} &=& 	 \left\{ 1 + \tanh \left[ \frac{2}{3} \alpha (N-N_f) \right]	+ \exp(-4 \alpha N)  \right\}  \nn
	&\times& \exp \left[ \frac{1}{2 \alpha}  \left(  -1 + e^{-4 \alpha N} -4 \alpha N \right)   \right] \nn
	&\times& \left\{    \frac{\cosh (2\alpha N_f/3)}{\cosh \left[ 2\alpha \left(N - N_f \right)/3 \right]}\right\}^{3/\alpha}
\end{eqnarray}
and
\begin{eqnarray}\label{Qcaso2}
	\frac{Q(N)}{M_P^4} &=&   \left\{ 1 - \tanh \left[ \frac{2}{3} \alpha (N-N_f) \right]	- \exp(-4 \alpha N)  \right\}  \nn
	&\times& \exp \left[ \frac{1}{2 \alpha}  \left(  -1 + e^{-4 \alpha N} -4 \alpha N \right)   \right] \nn
	&\times& \left\{    \frac{\cosh (2\alpha N_f/3)}{\cosh \left[ 2\alpha \left(N - N_f \right)/3 \right]}\right\}^{3/\alpha}.
\end{eqnarray}
Substituting Eqs. \eqref{rhocaso2} and \eqref{Qcaso2} into Eq.  \eqref{F1}, we can obtain $H(N)$.  The initial conditions chosen are $\rho(0) =\rho_0 =M_P^4$,  hence $Q(0)=M_P^4$. In other words, in this model we have selected natural initial conditions.  The  only parameter in this case is $\alpha$.  

Another important feature in this scenario is that  for certain values of $\alpha$, this case reproduces the main characteristics of the model originally introduced in \cite{aperez2021} (we will be more specific in the next section).  In such a model, the energy-momentum non-conservation is motivated by considering a fundamental granularity of the spacetime at Planckian scales. The diffusion term that arises in that model can be used to characterize an inflationary phase. However, in that same work, the primordial spectrum is generated by resorting to fluctuations of the Higgs scalar field during the inflationary regime within the semiclassical gravity framework.  If one chooses to drop the semiclassical gravity hypothesis and instead quantize simultaneously the metric and matter perturbations, as it is done in standard slow-roll inflation, then the present scenario includes the model of Ref. \cite{aperez2021}. Therefore, the parameterization \eqref{eps1caso2},  which led to  $\rho$ and $Q$ as in Eqs. \eqref{rhocaso2} and \eqref{Qcaso2} can be considered as a refinement of the model in Ref. \cite{aperez2021}

\subsection{Third Scenario}

In this scenario we present a method for reconstructing a particular $Q$ given a single field slow-roll potential with one parameter. In this manner, we can argue that any single field inflationary model of the slow-roll type can be mapped to an inflationary model in non-conservative UG for a specific $Q$.  The procedure is as follows.

For a given particular potential $V(\phi)$, with a single parameter ${\lambda}$, one calculates  $\phi$ as a function of the number of e-folds $N$ to the end of inflation; we denote such period of e-foldings as $\DN \equiv N_f - N$.  In particular, we need to solve the equation of motion for the homogeneous part of the field $\phi(t)$ in the slow-roll approximation together with Friedmann's equation. That is, after a change of variables $N(t)$, the equations to solve are: $3 H^2 \simeq  V/M_P^2$ and $ 3 H^2 \phi_{,N} \simeq - \partial_\phi V$. Those equations can be combined to yield
\begin{equation}\label{eqxx}
	\frac{d \phi}{d N} = -M_P^2 \frac{d \ln V}{d \phi},
\end{equation}
Denoting by $\mI$ the primitive
\begin{equation}\label{prim}
	\mI_{\lambda} (\phi) \equiv \int^\phi d\varphi \frac{V_{\lambda}(\varphi)}{\partial_\phi V_{\lambda} (\varphi)},
\end{equation}
equation \eqref{eqxx} can be solved
\begin{equation}
	N =  - \frac{1}{M_P^2} [ \mI_{\lambda}(\phi ) - \mI_{\lambda}(\phi_{\ini}) ]. 
\end{equation}	
So, we have
\begin{subequations}
	\begin{equation}
		N_f = - \frac{1}{M_P^2} [ \mI_{\lambda}(\phi_{\fin} ) - \mI_{\lambda}(\phi_{\ini}) ],
	\end{equation}
	\begin{equation}
		N = - \frac{1}{M_P^2} [ \mI_{\lambda}(\phi) - \mI_{\lambda}(\phi_{\ini}) ], 
	\end{equation}
\end{subequations}	
From the previous expressions, it follows that
\begin{equation}\label{phi*}
	\phi = \mI_{\lambda}^{-1} [ \mI_{\lambda}(\phi_{\fin}) + M_P^2 \DN     ].
\end{equation}

For a particular given potential $V_\lambda (\phi)$, equation \eqref{phi*} allow us to express the first HFF  in terms of $\lambda$ and $\DN$, i.e.  $\epsilon_1(\lambda,\DN)$. This is useful since we can use such an expression in Eqs. \eqref{rhosolNgen} and \eqref{QNgen}, which will lead us to the reconstruction of the diffusion term.  

The first step is to use  the equation of $\epsilon_1$ in terms of the potential $V$ and its derivative $\partial_\phi V$  (see \cite{terreroHFF,terreroHFF2}), namely
	\begin{equation}\label{PSR}
		\epsilon_1 (\lambda,\phi) \simeq \frac{M_P^2}{2} \left( \frac{\partial_\phi V}{V}\right)^2.
	\end{equation}
Inserting the explicit form of the potential as a function of $\phi$, i.e. $V_\lambda (\phi)$, into \eqref{PSR} results in  $\epsilon_1(\lambda,\phi)$. Subsequently, substituting solution  \eqref{phi*}, which is of the form $\phi(\DN)$, into \eqref{PSR}, one can finally obtain $\epsilon_1(\lambda,\DN)$. Recall that we can find the value of the field at the end of inflation by using the condition $\epsilon_1(\phi_{\fin}) \simeq 1$.  At this point, we are done with slow-roll inflation. 

With the calculated expression of the HFF at hand, $\epsilon_1(\lambda,\DN)$, we can reconstruct the corresponding diffusion term (as in the previous two scenarios) by using Eqs. \eqref{rhosolNgen},\eqref{QNgen}, and choosing $\rho_0 = \rho_\fin$, $N_0 = N_f$.  This is, we can reconstruct the corresponding inflationary phase in non-conservative UG based on a particular inflationary slow-roll model.  

As a concrete example for this scenario, we can focus on large field inflationary models characterized by a potential of the power law type, i.e.
\begin{equation}\label{potencial}
	V(\phi) = M^4 \left( \frac{\phi}{M_P} \right)^p.
\end{equation}
This example is also relatively easy to handle since all expressions involved, in the aforementioned method, can be obtained analytically. Following the previous procedure (from Eq. \eqref{eqxx} up to Eq. \eqref{PSR}), we obtain 
\begin{equation}\label{eps1caso3}
	\epsilon_1(N) = \frac{p}{4(N_f-N)+p},
\end{equation}
so 
\begin{equation}\label{eps2caso3}
	\epsilon_2(N) = \frac{1}{N_f-N+p/4}.
\end{equation}

In this case, using \eqref{rhosolNgen},\eqref{QNgen},  we obtain
\begin{equation}\label{rhocaso3}
	\rho(N) = \rho_\fin \left[ \frac{p}{4(N_f-N)+p} \right]^{1-p/2}
\end{equation}
and
\begin{equation}\label{Qcaso3}
	Q(N) = \rho_\fin \left[2 - \frac{p}{4(N_f-N)+p} \right] \left[ \frac{p}{4(N_f-N)+p}  \right]^{-p/2}
\end{equation}
Lastly, from Friedamnn's equation, we have
\begin{equation}\label{Hcaso3}
	H^2(N) = \frac{2 \rho_\fin}{3 M_P^2} \left[ \frac{4}{p} (N_f-N) + 1   \right]^{p/2}.
\end{equation}
The two parameters in this case are:  $p$  and $\rho_\fin$.

\section{Three inflationary scenarios: Observational Analysis}\label{Sec_observations}

In order to test the previous three scenarios with observational data, we proceed to
run publicly available numerical codes: {\tt CAMB} \cite{CAMB} and
{\tt COSMOMC} \cite{COSMOMC}. The first computes CMB anisotropies by integrating the differential coupled equations describing the  primordial universe, while
the second implements Markov--Monte Carlo chains to estimate a set of cosmological parameters.

The standard $\Lambda$CDM cosmological model was used in order to establish a baseline reference to distinguish whether or not our proposed scenarios fit the latest observational data. The dataset implemented consisted of  \textit{Planck 2018} TT,TE,EE + lowEB + lensing + BK15, which combines latest \textit{Planck} temperature and polarization data \cite{Planck18c} with BICEP2/Keck 2015 release \cite{KECKPLANCK}. We found the aforementioned combination of datasets to be the most adequate for our purpose.

The usual set of standard cosmological parameters was allowed to vary,
that is to say $\Omega_b h^2$ (baryon density today), $\Omega_c h^2$
(cold dark matter density today), $\tau$ (Thomson scattering optical
depth due to reionization), $100\Theta_{MC}$ (100 $\times$
approximation to $r_s / D_A$), ${\rm{ln}}(10^{10} A_s)$ (log power of
the primordial curvature perturbations) and $n_s$ (scalar spectrum
power--law index), plus $r$ (tensor power spectrum amplitude). As for the pivot scale, we take the standard value $k_0 = 0.05 \: \rm{Mpc}^{-1}$. 

The runs performed with the previous input allowed us to create marginalized joint 68\% and 95\% confidence level regions for the cosmological parameters of the $\Lambda$CDM model.  In particular, we focused on the inflationary parameters ${\rm{ln}}(10^{10} A_s)$, $n_s$ and $r$. For each confidence region,   we overlay the theoretical predictions of each model to evaluate their compatibility with the data. An extra feature added to the majority of our plots is a star point $\star$, which singles out a specific set of values for the parameters characterizing each scenario.  The $\star$ symbol represents a reference model that ensures the model's predictions are consistent with observational data for those specific parameter values. 

In the next subsections we will analyze in detail for each proposed scenario the behavior of the functions $Q(N)$ and $\rho(N)$.  Also, employing the parameter estimation contours built by the above procedure, together with specific theoretical plots,  we will analyze the feasibility of  each inflationary UG model  by considering the variation of the corresponding relevant parameter(s).

\subsection{First scenario}\label{first}
\begin{figure*}
	\centering
	\includegraphics[angle=270,width=0.5\textwidth]{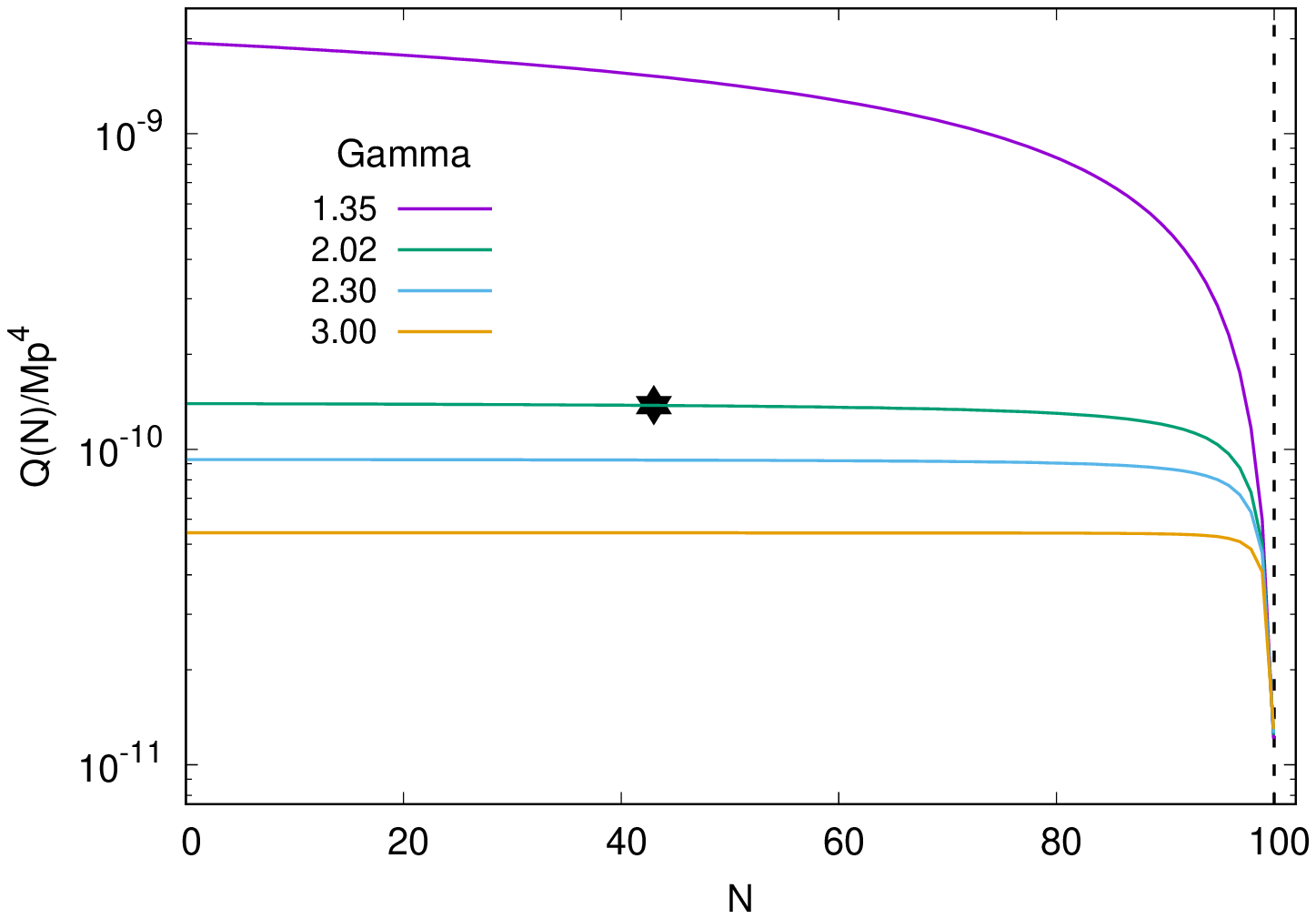}~%
	\includegraphics[angle=270,width=0.5\textwidth]{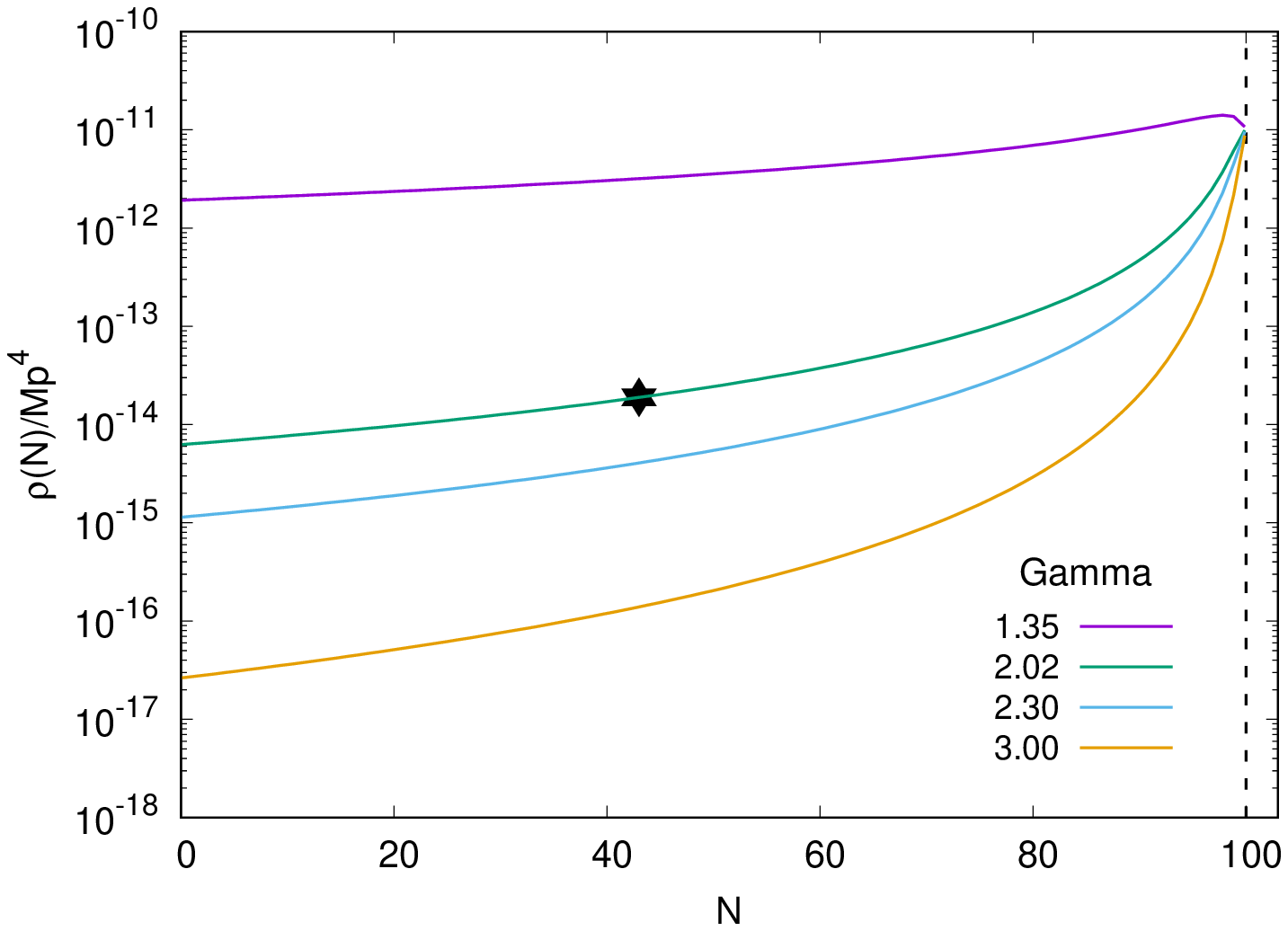}  
	\caption{$Q(N)/M_P^4$ and $\rho(N)/M_P^4$ (Eqs. \eqref{Qcaso1} and  Eq. \eqref{rhocaso1} respectively) are plotted for different values of $\gamma$ and a fixed value for $\rho_\fin=10^{-11} M_P^4$. The star marks  the specific values  $N=43$ and $\gamma=2.02$, as a reference which is compatible with the observational data. Note that total number of e-folds that inflation lasts is chosen to be $N_f = 100$, shown as a vertical dashed line.}
	\label{fig:gamma_lineplot}
\end{figure*}

The functions $Q(N)/M_P^4$ and $\rho(N)/M_P^4$, given by Eqs. \eqref{Qcaso1} and \eqref{rhocaso1}, are plotted in Fig. \ref{fig:gamma_lineplot}. The total number of e-folds of inflation here is taken to be $N_f = 100$.  The energy density at end of inflation is chosen to be fixed at $\rho_\fin = 10^{-11} M_P^4$. We have chosen that scale because it is a reasonable energy scale for the beginning of the radiation dominated epoch, $\sim 10^{-15}$ GeV.  
The $\gamma$  parameter in this scenario is allowed to vary; the different color lines refer to a set of chosen values, characterizing a range of feasible values. 

\begin{figure*}
	\centering
	\includegraphics[width=0.45\textwidth]{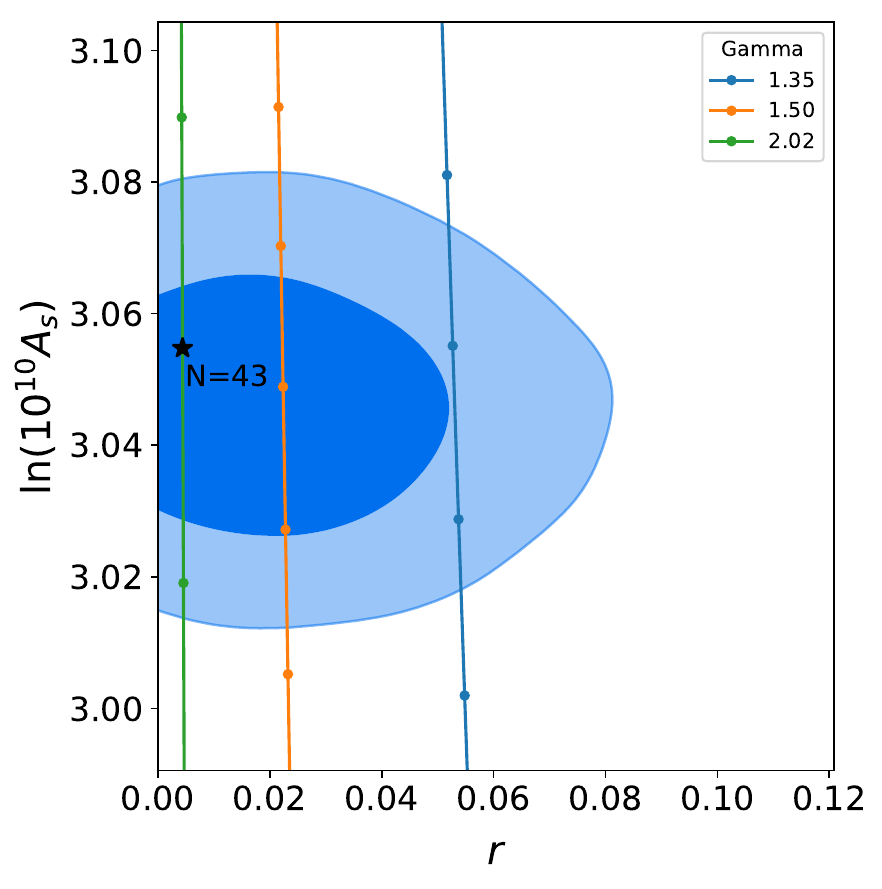}~\hspace{2em}%
	\includegraphics[width=0.45\textwidth]{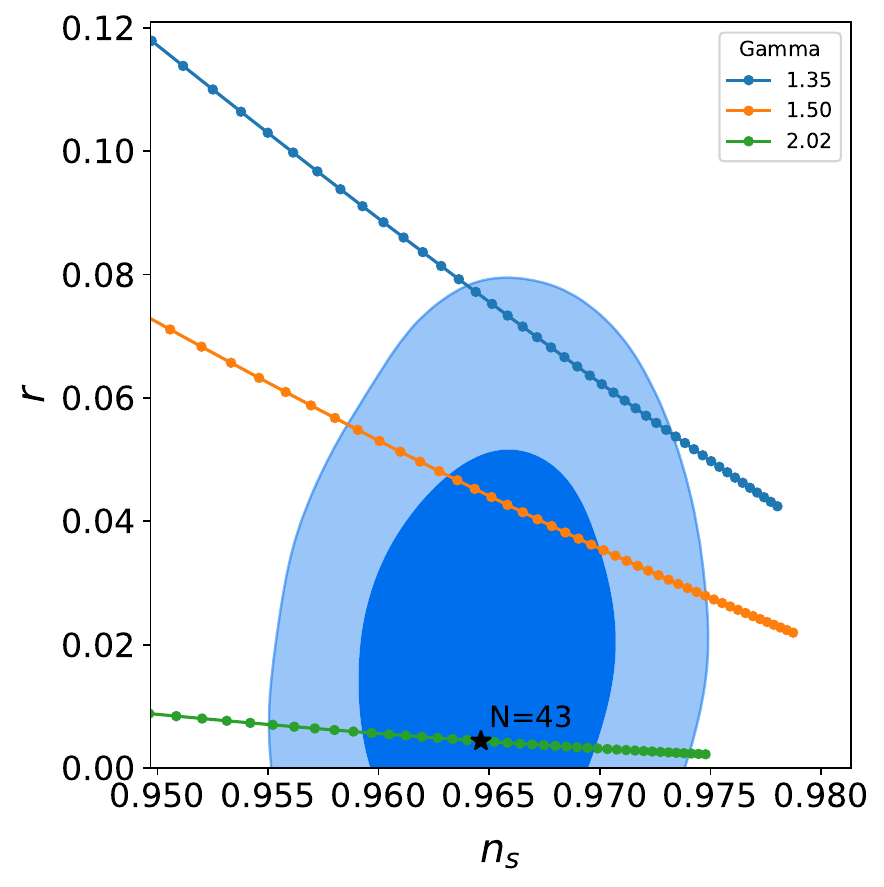}
	\caption{Confidence regions for the first scenario, comparing parameter values  of interest. Different values of $\gamma$ are tested, proving good potential to fit the data. The star point marks  $N=43$ and $\gamma = 2.02$, which we present as a reference.}
	\label{fig:contornos1y2_gamma}
\end{figure*}

\begin{figure*}
	\centering
	\includegraphics[width=0.45\textwidth]{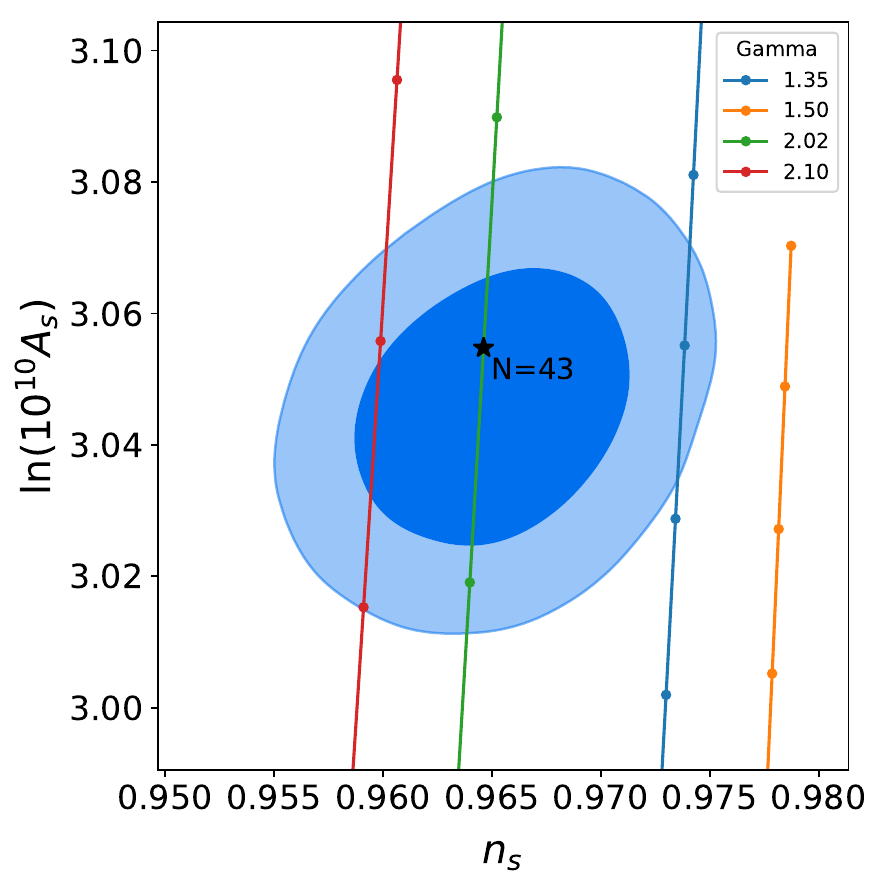}  
	\caption{Scalar spectral index vs. scalar amplitude at $68\%$ and $95\%$ confidence levels. In this case, some of the positively evaluated values for $\gamma$ in the previous figure are left outside these regions, restringing more the acceptable interval for the free   parameter. The particular choice of $\gamma = 2.02$ proves to pass successfully all the tests.  }
	\label{fig:contorno3_gamma}
\end{figure*}

In Figs. \ref{fig:contornos1y2_gamma} and \ref{fig:contorno3_gamma}, we show a  comparison between the predictions of the theoretical model and the parameter likelihood contours characterizing inflation, i.e.   $r$ vs ${\rm{ln}}(10^{10} A_s)$, $n_s$ vs $r$ and $n_s$ vs ${\rm{ln}}(10^{10} A_s)$. As can be seen, not any value shown in Fig. \ref{fig:gamma_lineplot} is consistent with the data. This separation contributes to the predictability of the model. Namely, despite the fact that a range of $\gamma$ values are allowed in the theoretical approach, not all of them are eligible by the observational data.

Continuous lines show the prediction for some possible values of $\gamma$, while the dots depict different e-folds. The values taken as a reference, identified by  the $\star$ symbol, are inside the $68\%$ CL region, these are $N=43$  and $\gamma = 2.02$.  Therefore, those values yield theoretical predictions that are consistent with observations. On the other hand, $\gamma = 1.5$ enters the $68\%$ CL regions shown in Fig. \ref{fig:contornos1y2_gamma},  but is excluded by the data at a $95\%$ CL when comparing $n_s$ and ${\rm{ln}}(10^{10} A_s)$ , see Fig.\ref{fig:contorno3_gamma}.

\subsection{Second scenario}\label{second}

\begin{figure*}
	\centering
	\includegraphics[angle=270,width=0.5\textwidth]{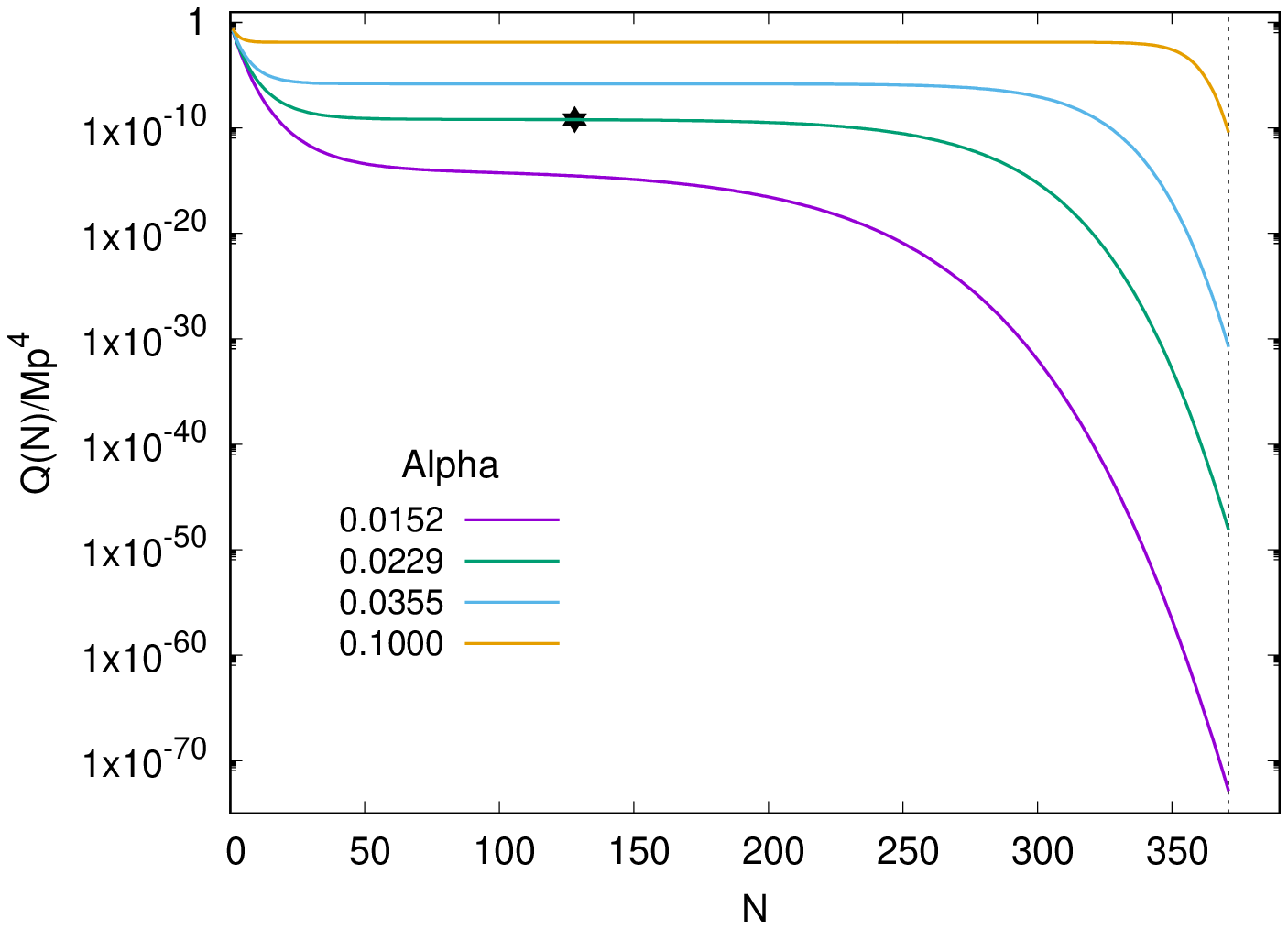}~%
	\includegraphics[angle=270,width=0.5\textwidth]{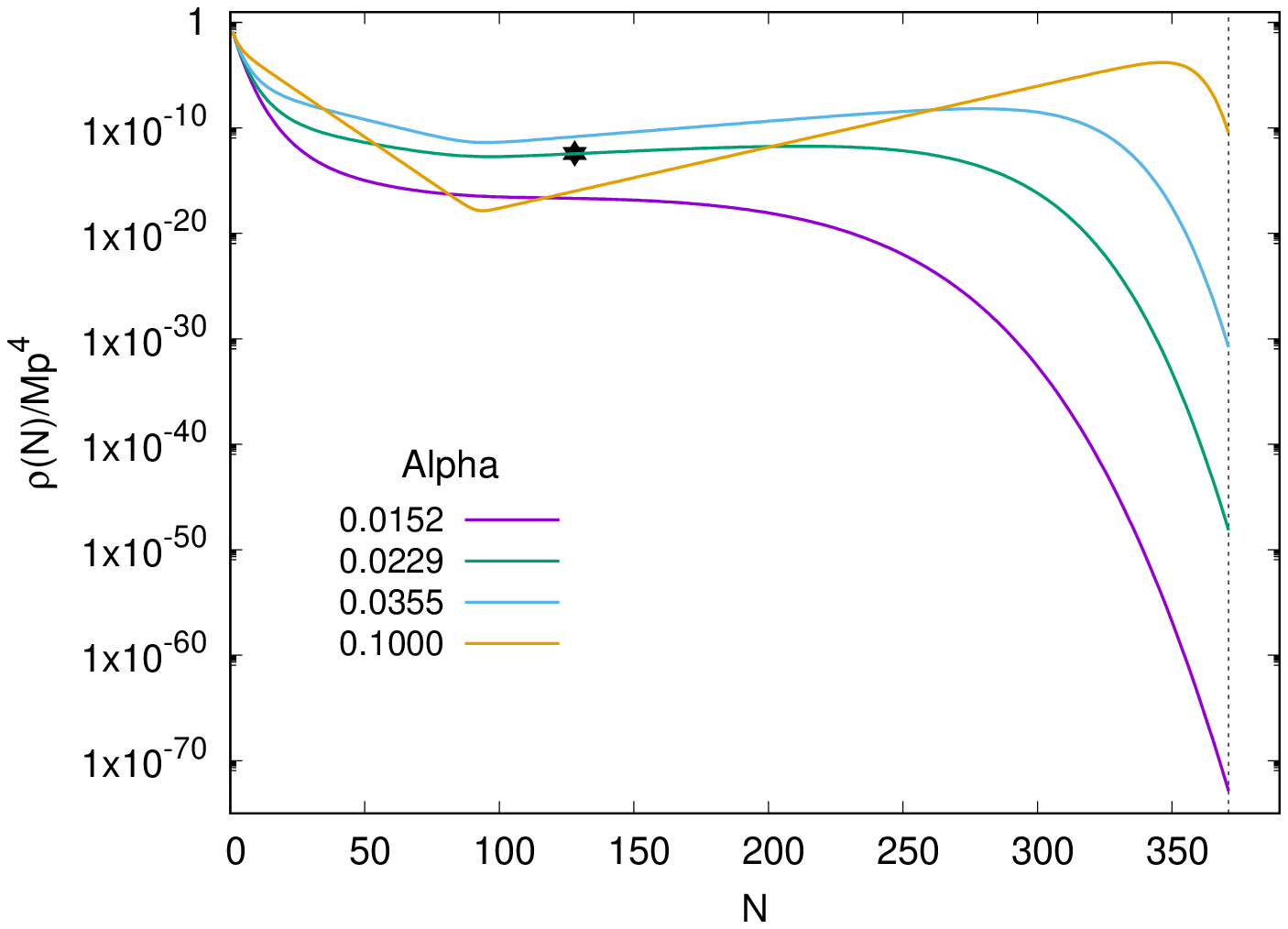}
	\caption{The functions $Q(N)/M_P^4$ and $\rho(N)/M_P^4$ [Eqs.~\eqref{Qcaso2} and \eqref{rhocaso2} respectively] are plotted assuming different values of $\alpha$. A star depicts the specific value for $N=128$ and $\alpha = 0.0229$. The value $\alpha=0.1$ reproduces the AP model. }
\label{fig:alpha_lineplot}
\end{figure*}

In the second scenario, Eqs.~\eqref{Qcaso2} and \eqref{rhocaso2} are plotted in Fig.~\ref{fig:alpha_lineplot} corresponding to the functions $Q(N)/M_P^4$ and $\rho(N)/M_P^4$ respectively. In this case, we have assumed $N_f=371$, which is depicted as a vertical dashed line. Several values of the free parameter $\alpha$ are shown. The value $\alpha=0.1$ has particular interest as it reproduces the same dynamical behavior of $\rho$ and $Q$ of Ref. \cite{aperez2021}, from now on we refer to that model as the AP model.  

At first glance, some value of $\alpha$  between $[10^{-2},10^{-1}]$ seems to be a good potential choice. Clearly, this indicates that we can consider such a range as a \textit{prior} probability for future  parameter estimation with Markov chains in order to find the best fit--to--data.

\begin{figure*}
	\centering
	\includegraphics[width=0.45\textwidth]{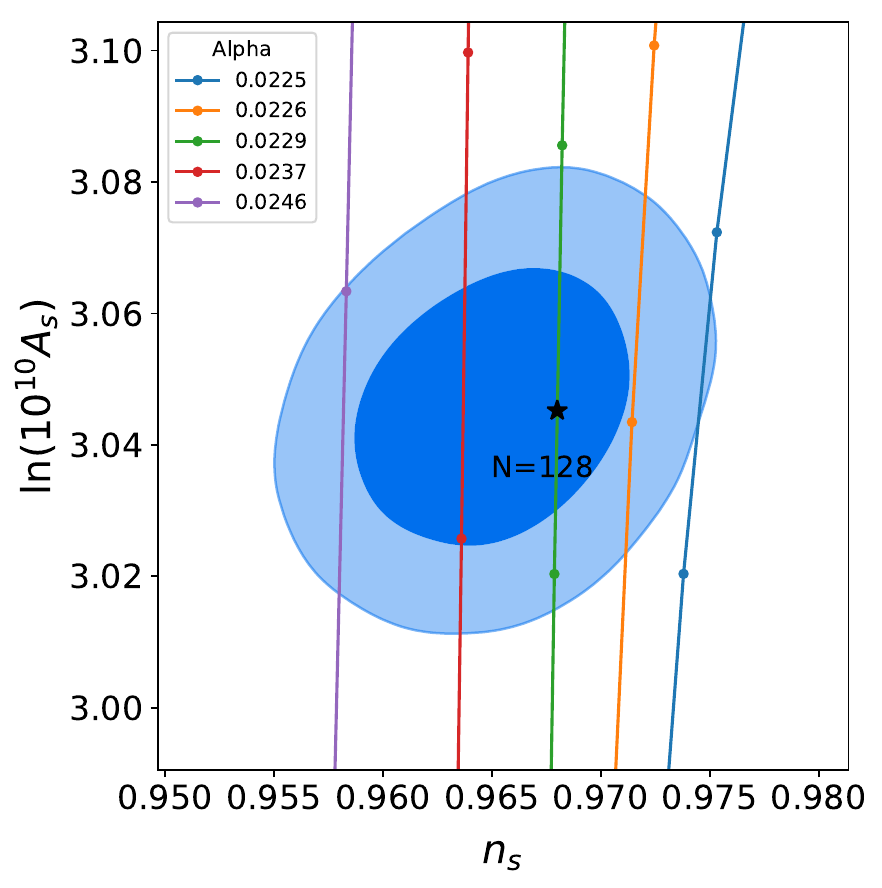}~\hspace{2em}
	\includegraphics[width=0.45\textwidth]{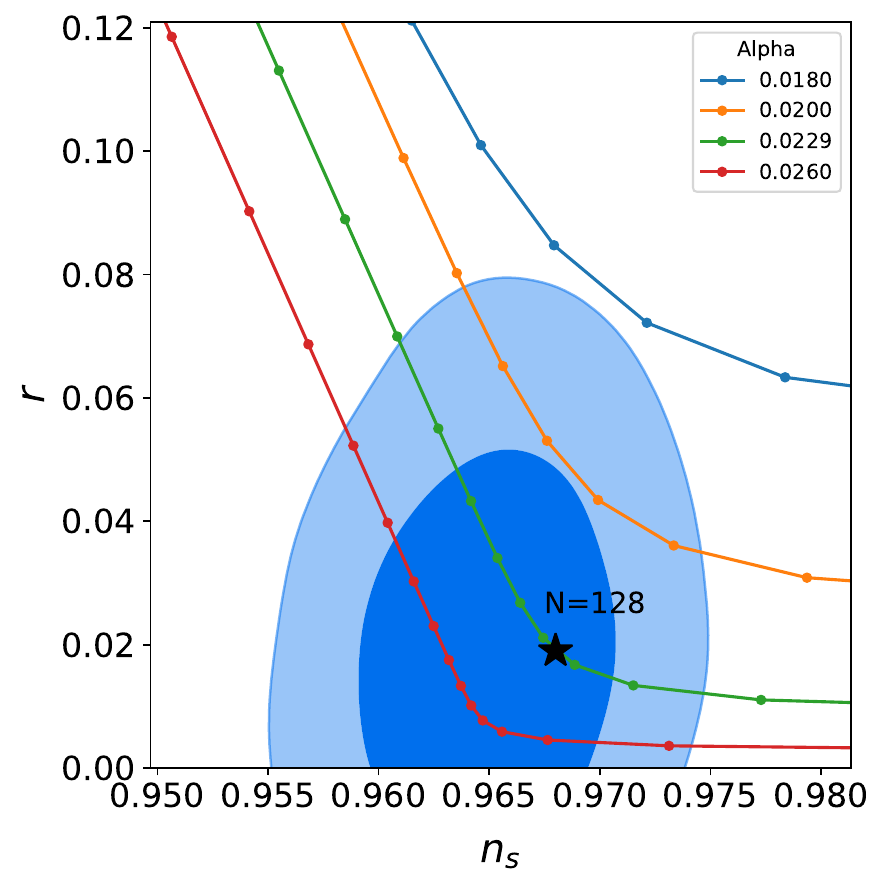}
	\caption{Blue contours show \textit{Planck}-- BICEP2/Keck marginalized joint 68\% and 95\% CL regions for the inflation parameters $n_s$, ${\rm{ln}}(10^{10} A_s)$ and $r$ in the $\Lambda$CDM model. Color lines denote different values of the free parameter $\alpha$ for the second inflationary scenario in UG.}
	\label{fig:contornos1y2_alpha}
\end{figure*}

\begin{figure*}
	\centering
	\includegraphics[width=0.45\textwidth]{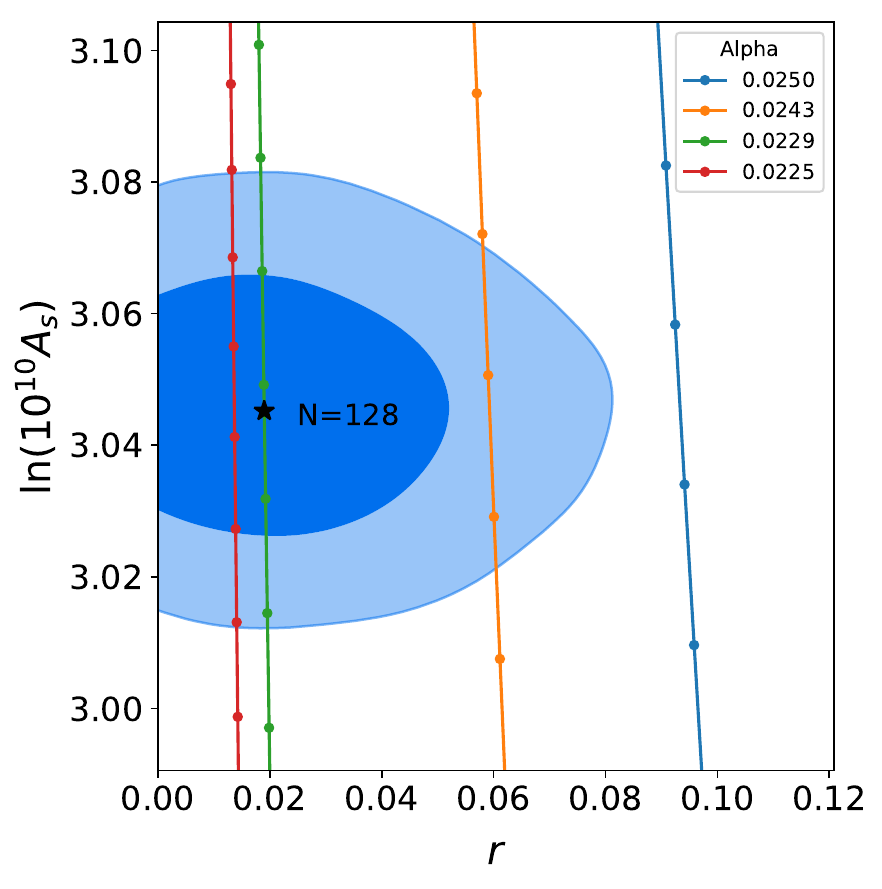}
	\caption{Stringent restrictions for the value of $\alpha$ arise when tensor-to-scalar ratio and scalar spectral index regions are built from the latest observational data.  This contour progressively reduces the acceptable margin of validity for the free parameter in the second scenario.  Also, it shows that $\alpha = 0.0229$ is a prediction consistent with all the data, hence the $\star$ symbol.}
	\label{fig:contorno3_alpha}
\end{figure*}

Once again, the star point $\star$ singles out the values of $N$ and $\alpha$  that suit observations for this scenario.  The contours regions $n_s$ vs ${\rm{ln}}(10^{10} A_s)$ and $n_s$ vs $r$ are shown in   Fig. \ref{fig:contornos1y2_alpha}. There we can observe that some values for  $\alpha$ will be excluded by the observational data at a $95\%$ CL, but some others will prevail, giving predictability to the model. Moreover, the contour $r$ vs~${\rm{ln}}(10^{10} A_s)$ shown in Fig. \ref{fig:contorno3_alpha} will result in stricter exclusion values for the model's parameters.   In particular, the value $\alpha=0.1$, corresponding to the AP model (without semiclassical gravity) is ruled out by the data (it does not even appear close to the confidence regions).  However, we note that the value $\alpha=0.0229$ can be considered as  an excellent guess as a starting point for a {\tt COSMOMC} run. The ``drawback'' of this scenario is that it requires ``too much'' inflation, i.e. $N_f \simeq 370$.

\subsection{Third scenario}\label{third}

As we have mentioned, this last scenario involves actually many models since it maps any single-field slow-roll model to a specific form of the diffusion term $Q$ that can generate an inflationary expansion. Therefore, the results for this scenario are already known because in principle the predictions obtained are the same as the ones from slow-roll inflation. 

To test this assumption, we choose the power law model of single-field slow-roll inflation, i.e. the one characterized by the potential in Eq. \eqref{potencial}.  Latest results from \textit{Planck} collaboration have already ruled out practically any model of the power law type \cite{Planck18b}.  Thus, we know in advance what we should expect in this case.  

\begin{figure*}
	\centering
	\includegraphics[angle=270,width=0.5\textwidth]{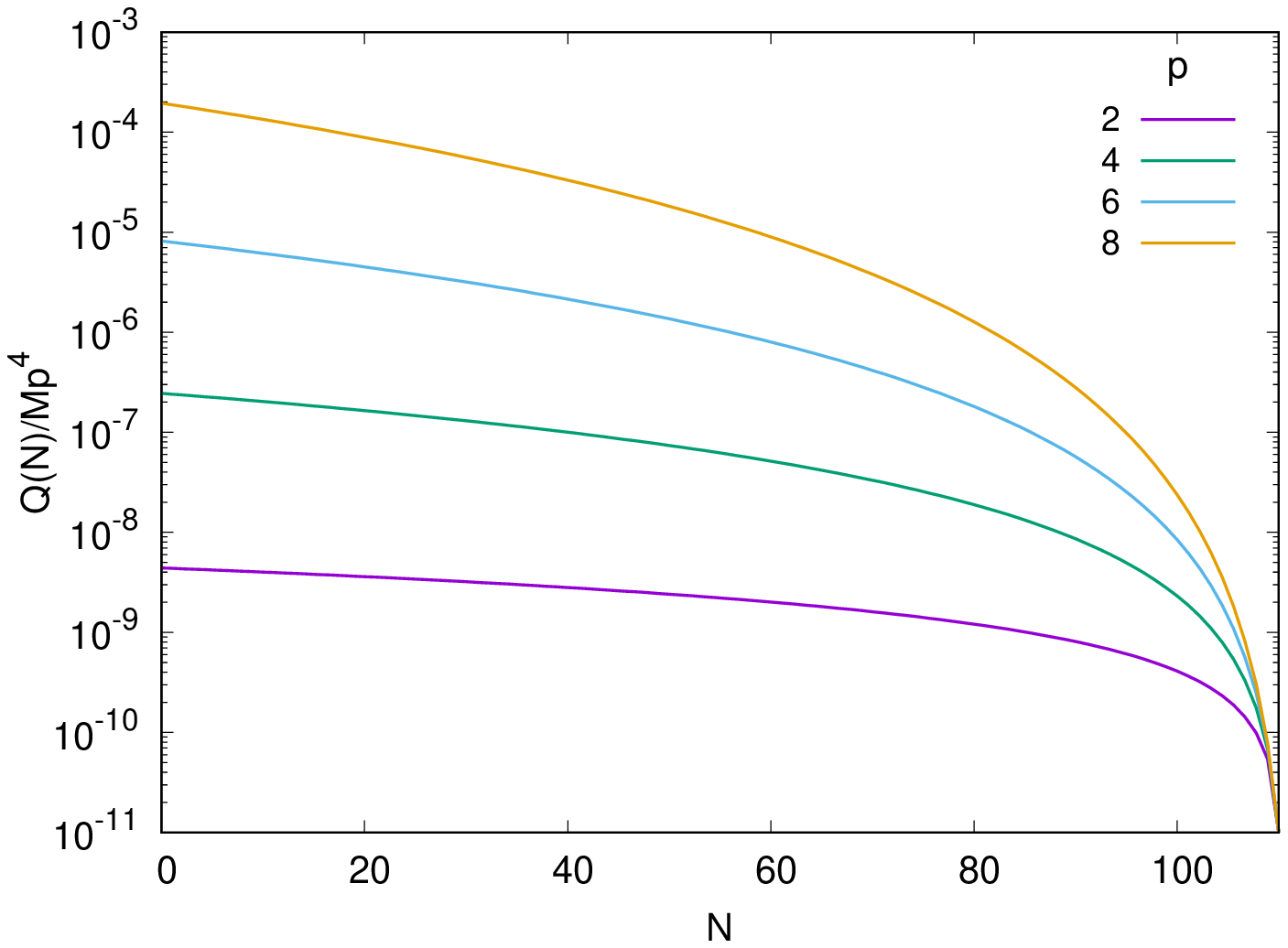}~
	\includegraphics[angle=270,width=0.5\textwidth]{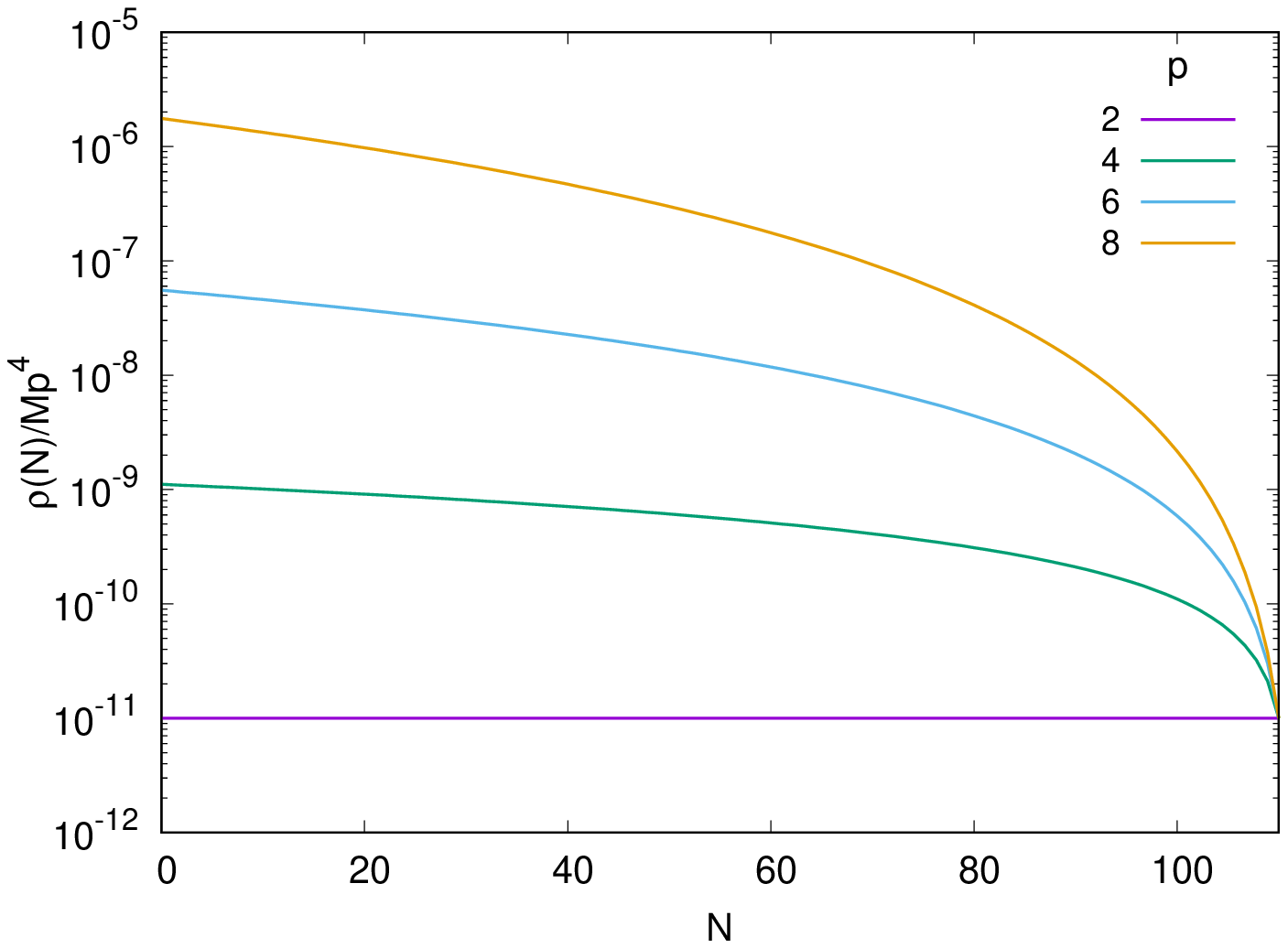}
	\caption{ Plots of the functions $Q(N)/M_P^4$ and $\rho(N)/M_P^4$ [Eqs.~\eqref{Qcaso3} and \eqref{rhocaso3} respectively] for the third scenario. We vary the free parameter $p$  between $2$ and $8$ to illustrate the corresponding behavior in this case.}
	\label{fig:p_lineplot}
\end{figure*}

We fix the energy density at the end of inflation $\rho_\fin = 10^{-11} M_P^4$ and $N_f = 100$ for the exact same reason as in the first scenario.  In Fig. \ref{fig:p_lineplot}, we plot the functions $Q(N)/M_P^4$ and $\rho(N)/M_P^4$ with $2<p<8$. This time no $\star$ point is drawn because there are no suitable values $N$ and $p$ consistent with the data.

\begin{figure*}
	\centering
	\includegraphics[width=0.45\textwidth]{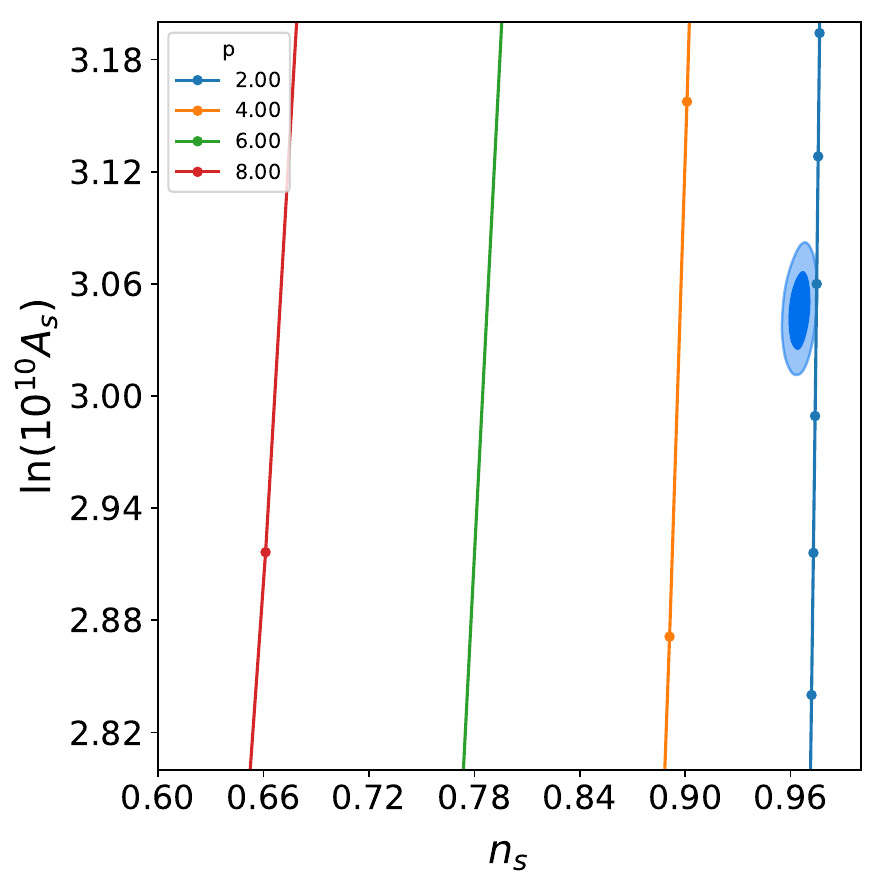}~\hfill%
	\includegraphics[width=0.45\textwidth]{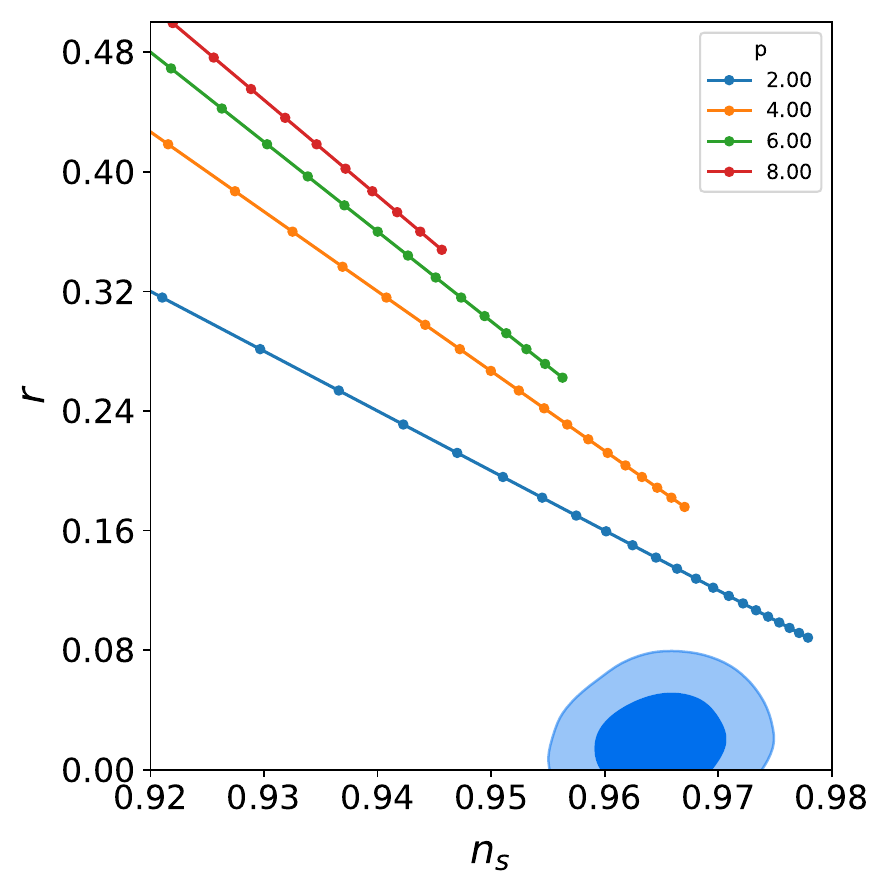}
	\caption{Contours of  the scalar spectral index vs.~scalar  amplitude/tensor-to-scalar ratio show that no value of the free parameter $p$ fall inside any confidence  region. Thus, the UG inflationary model,  inspired by  the single-field  power law potential, is ruled out by the data.}
	\label{fig:contornos1y2_p}
\end{figure*}

\begin{figure*}
	\centering
	\includegraphics[width=0.42\textwidth]{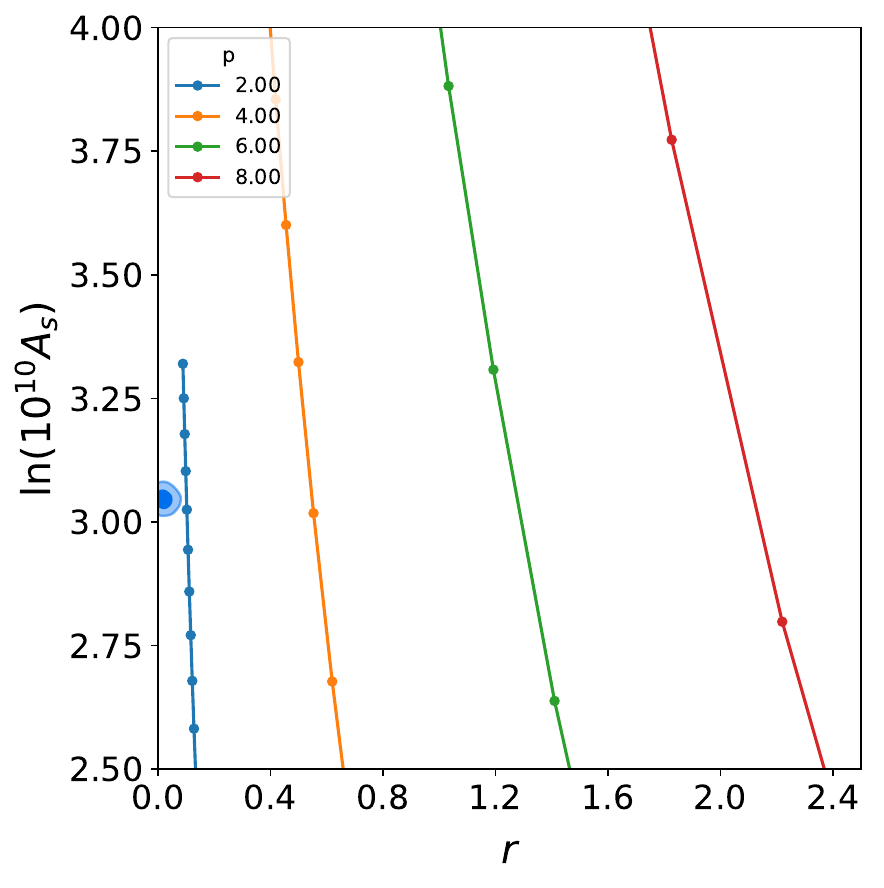}  
	\caption{The lowest possible value $p=2$ is barely close to the $95\%$ confidence level region in the ${\rm{ln}}(10^{10} A_s)$ vs $r$ contour. Theoretical predictions become less compatible with the observational data as $p$ increases. }
	\label{fig:contorno3_p}
\end{figure*}

We proceed to perform the same tests as before for this case. Figs.~\ref{fig:contornos1y2_p} and \ref{fig:contorno3_p} show 68\% and 95\% CL regions for observational data, comparing together $n_s$, $r$ and ${\rm{ln}}(10^{10} A_s)$. No possible value of $p$ is found such that the predicted quantities fall inside the confidence regions. This  indicates that an inflationary expansion in UG with a  diffusion term that can be mapped  to  slow-roll inflation  of the power law type is ruled out, as expected,  by the  latest cosmological observations.

\section{Conclusions}\label{Sec_conclusions}

The structure of unimodular gravity (UG) allows for a possible non-conservation of the canonical energy-momentum tensor, resulting in a so called diffusion term $Q$.   This diffusion term could be responsible for generating a realistic inflationary phase \cite{gabrielUG}, perhaps due to a fundamental granularity of the spacetime \cite{aperez2021}.  That is, in this approach, there is no need to postulate the existence of the inflaton  to produce an inflationary expansion in the early universe.  Moreover, the primordial inhomogeneities arise from the inhomogeneous part of standard hydrodynamical matter, modeled as a single ultra-relativistic fluid, i.e. pure radiation. 

In this article,  we have presented a phenomenological analysis involving three different inflationary scenarios in non-conservative UG, each with promising theoretical perspectives.  The inflationary scenarios considered were characterized through the Hubble Flow Functions (HFF).  The HFF allowed us to reconstruct the corresponding diffusion function $Q(N)$, which closed the set of cosmological equations.  Consequently, we have obtained the corresponding predictions for the scalar amplitude $A_s$, the scalar spectral index $n_s$ and the 
tensor-to-scalar ratio $r$.   The three scenarios were challenged to fit observational data in order to be deemed eligible from a realistic point of view. Specifically, the conjunct Planck temperature/polarization and Bicep/KECK collaboration results were used for this purpose. Based on these data, marginalized joint 68\% and 95\% confidence level regions were constructed for the cosmological parameters $r$, $n_s$, and $\ln (10^{10} A_s)$.

 In the first scenario, the HFF were  parameterized in the most simple manner  using a single parameter $\gamma$, Eqs.  \eqref{eps1caso1} and \eqref{eps2caso1}.  Setting the total inflation period to $N_f = 100$ e-folds, and the energy density at the end of inflation to $\rho_\fin = 10^{-11} M_P^4$, we were able to progressively restrict a feasible value for $\gamma$ while discriminating others. In particular, $\gamma \simeq 2$ would seem to provide a good fit to the data, see Figs.  \ref{fig:contornos1y2_gamma}, \ref{fig:contorno3_gamma}.  Thus, this scenario is a potentially strong candidate as an alternative to single-field slow-roll inflation.

The second scenario was motivated by Ref. \cite{gabrielUG}, which explored the idea that the diffusion term  could account  simultaneously for the  inflationary period and the current value of the cosmological constant, i.e. using the same $Q$.  Another interesting motivation in this case was the natural initial conditions considered, these are:  $\rho_0 \simeq Q_0 \simeq M_P^4$ at the beginning of inflation. The corresponding parameterization  in this scenario  involved a single parameter $\alpha$.     Additionally, such parameterization includes, as the particular case $\alpha = 0.1$,  an equivalent model as the one presented originally in Ref. \cite{aperez2021}, which we refer to as the AP model.  Our results indicate that only values very close to $\alpha = 0.022$ appear to be a suitable match for the data, see Figs.  \ref{fig:contornos1y2_alpha}, \ref{fig:contorno3_alpha}. However, one possible shortcoming of this scenario is that it requires $N_f \simeq 370$ e-folds of total duration of inflation.  While there is no upper bound for how much inflation should last that is  imposed by observations or theoretically, the number $N_f \simeq 370$ is almost four times more than what would normally be expected.  Further analysis is required for this scenario, especially if the initial intent was to account for the present value of the cosmological constant.  On the other hand, the value $\alpha = 0.1$, corresponding to the AP model, is not compatible with the data. Nonetheless, the ruled out AP model is not exactly the same as the one in Ref. \cite{aperez2021}. The former is based on the standard  procedure in which both   the metric and matter perturbations are quantized, while the latter was developed using the semiclassical gravity framework,  where only the matter fields are subjected to quantization.

The third scenario actually involves many different models. Specifically, we have shown how to map any single-field slow-roll inflation model, characterized by its potential, to an inflation model in non-conservative UG. Therefore, the predictions of any slow-roll inflation model can be reproduced in UG,  where the inflationary expansion  is driven by a particular diffusion function $Q$.  As a practical example,  we considered a power law type of potential and found its corresponding diffusion term.  Our  analysis is consistent with what is already known about this particular model \cite{jmartinpotentials,Planck18b}.  Namely, it is not compatible with the latest observational data, see Figs. \ref{fig:contornos1y2_p}, \ref{fig:contorno3_p}.

We conclude  that further research is necessary to establish our proposal as a solid alternative to traditional single-field slow-roll inflation. In particular, it is required to formally develop the quantum field theory of the matter fields involved, e.g. by considering the Higgs field (and its fluctuations) as in \cite{aperez2021}.  Another open aspect, is to clearly explain the generation of primordial inhomogeneities from the microphysics which also produces the $Q$ term.  We hope to address those (and other possible) issues in future works.

\section*{Acknowledgements}
G.L. is supported by CONICET (Argentina). G.L and M.P.P acknowledge support from the following project grants: Universidad Nacional de La Plata I+D  G175 and  PIP 11220200100729CO  CONICET (Argentina). M.P.P thanks Facultad de Ciencias Astron\'omicas y Geof\'{\i}sicas UNLP, as this work was mainly done under \textit{Programa de Retenci\'on de Recursos Humanos}.
We are  especially grateful to the anonymous referee for a helpful review. Their comments and suggestions have led to significant improvements in the presentation of the material in this manuscript.

\section*{Data Availability}
Observational constraints used in this article were obtained
using {\tt Plik v3.1} likelihood software available at Planck Legacy Archive
\href{http://pla.esac.esa.int}{http://pla.esac.esa.int}, together with {\tt COSMOMC}, {\tt CAMB} and {\tt GetDist} codes available at \href{https://cosmologist.info}{https://cosmologist.info}. Markov-Monte Carlo chains underlying this article may be available upon request to the corresponding author.

\bibliography{bibliografia}
\bibliographystyle{apsrev}
\label{lastpage}
\end{document}